\documentclass[prl,twocolumn]{revtex4}

\usepackage{amsmath,amssymb,graphicx,color,bm,soul}
\usepackage{latexsym}
\usepackage{dcolumn}
\usepackage{epsf}
\usepackage{float}

\usepackage{tabularx}

\begin{document}

\title{Floquet quantum simulation with superconducting qubits}

\author{Oleksandr Kyriienko}
\affiliation{The Niels Bohr Institute, University of Copenhagen, Blegdamsvej 17, DK-2100 Copenhagen, Denmark}

\author{Anders S. S{\o}rensen}
\affiliation{The Niels Bohr Institute, University of Copenhagen, Blegdamsvej 17, DK-2100 Copenhagen, Denmark}

\date{\today}

\begin{abstract}
We propose a quantum algorithm for simulating spin models based on periodic modulation of transmon qubits.  Using Floquet theory we derive an effective time-averaged Hamiltonian, which is of the general XYZ class, different from the isotropic XY Hamiltonian typically realised by the physical setup. As an example, we provide a simple recipe to construct a transverse Ising Hamiltonian in the Floquet basis. For a 1D system we demonstrate numerically the dynamical simulation of the transverse Ising Hamiltonian and quantum annealing to its ground state. We benchmark the Floquet approach with a digital simulation procedure, and demonstrate that it is advantageous for limited resources and finite anharmonicity of the transmons. The described protocol can serve as a simple yet reliable path towards configurable quantum simulators with currently existing superconducting chips.
\end{abstract}

\maketitle

Quantum simulation generally relies on exploiting a controllable quantum system to imitate a complex quantum system of interest \cite{Lloyd1996}. It provides an efficient way to solve classically inaccessible problems of material science \cite{Wecker2015} and quantum chemistry \cite{Wecker2014,Reiher2016}. Ultimately quantum simulation may give access to properties of complex quantum systems with exponential speed-up as compared to classical algorithms \cite{NielsenChuang,Georgescu2014}.
Superconducting circuits (SC) \cite{Wallraff2004,Devoret2013} have recently emerged as one of the prime candidates for realizing full scale quantum computers, with operations close to the fault tolerant threshold \cite{Barends2014,Kelly2015,Riste2015,Corcoles2015,Ofek2016,Reed2012} and the possibility of scaling these advances to larger systems \cite{Bejanin2016}. From the simulation point of view, various protocols were proposed and realized experimentally \cite{Houck2012}, including simulation of spin systems \cite{LasHeras2014,Salathe2015,Barends2016}, fermionic models \cite{Barends2015}, quantum chemistry \cite{OMalley2015}, thermalization \cite{Neill2016}, synthetic magnetic fields \cite{Roushan2016}, ultrastrong coupling \cite{Ballester2012,Langford2016}, and gauge field theories \cite{Mezzacapo2015}. To demonstrate the full potential of quantum simulation with SC, however, there is a need for protocols which can outperform classical protocols for realistic near term medium size systems.

Typically protocols for quantum simulation can be divided into \emph{digital} and \emph{analog} (or \emph{emulator}) types. While these techniques are similar, they exploit different methods to achieve a quantum speed-up. 
The digital approach relies on discretizing the Hamiltonian evolution using a set of quantum gates. A protocol for simulating an arbitrary unitary $\hat{U}(t) = \exp(-i\hat{\mathcal{H}}t)$ governed by a Hamiltonian $\hat{\mathcal{H}}$ not available in a physical setup, exploits the sequential implementation of the available unitaries $\hat{U}_{\ell}$ represented by gates. The corresponding unitary of a single digital step $j$ of duration $\delta t$ can be constructed as $\hat{U}_{j} (\delta t) = \prod_{\ell} \hat{U}_{\ell}$. This string of unitaries can be recast in terms of Hamiltonians $\hat{U}_{j} (\delta t) = \prod_{m} e^{-i\hat{\mathcal{H}}_m \delta t}$. An implementation of $N_{\mathrm{Tr}} \rightarrow \infty$ of these \emph{Trotter} steps combines into a unitary $\hat{U}(t) = \lim_{N_{\mathrm{Tr}}\rightarrow \infty}\hat{U}_j(\delta t)^{N_{\mathrm{Tr}}} \approx e^{-i\hat{\mathcal{H}} t}$, where $\hat{H} = \sum_{m}\hat{\mathcal{H}}_m$ and $t = \delta t N_{\mathrm{Tr}}$ \cite{NielsenChuang}.
Given a universal set of gates any required Hamiltonian can in principle be simulated. This poses the challenge of implementing several quantum gates for successful simulation, leading to errors if they have limited fidelity. This issue can possibly be overcome by quantum error correction but this requires substantial overhead in resources. The digital approach is widely used for quantum simulation with SC \cite{LasHeras2014,Salathe2015,Barends2016,Barends2015,OMalley2015,Ballester2012,Langford2016}, as it is tunable and does not require changing sample layout to simulate different models.
\begin{figure}
\includegraphics[width=1.0\linewidth]{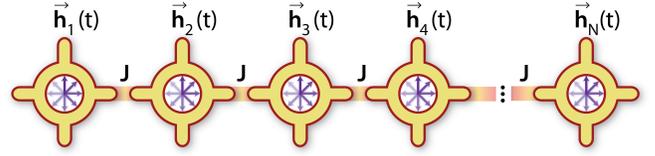}
\caption{\textbf{Sketch of the system}. A chain of superconducting transmon-type qubits coupled through isotropic XY coupling $J$. Each qubit is subject to a periodically modulated effective magnetic field $\mathbf{h}_j(t)$.}
\label{fig:sketch}
\end{figure}

Analog quantum simulation relies on the actual physical implementation of the required Hamiltonian, corresponding to the emulation of a targeted real system. This was realised on various platforms, including cold atoms in optical lattices \cite{Bloch2012,Simon2011} and trapped ions \cite{Kim2010,Cenko2015}. By exploiting continuous time dynamics, Trotterization errors are minimized and analog protocols can therefore have superior simulation fidelities compared to digital approaches. However, they are restricted to the types of Hamiltonian physically realizable in the system. 
In particular, this is the case for SC quantum systems. Current highly coherent chains of transmons, sketched in Fig. \ref{fig:sketch}, are limited to nearest neighbour flip-flop type of interqubit interaction provided by the capacitive coupling between them. Thus, they are confined to simulate isotropic XY type spin-1/2 model \cite{Wendin2016}. The accessible models can be enriched by implementing different connections between the qubits to engineer various nonlinear couplings \cite{Neumeier2013,Dumur2015,Sameti2016}, or allowing for modulation of interqubit interactions to break the rotating wave approximation \cite{Bertet2006,Mezzacapo2014,Kapit2015}. However, this adds extra complications to the experiments and potentially introduces additional errors.

Here, we propose an alternative analog-like simulation strategy, which can be performed without modifications of the system. It is based on using a Floquet basis to perform quantum simulation with superconducting circuits, and can be extended to ground state preparation via quantum annealing. The idea relies on the time-dependent modulation of the Hamiltonian $\hat{\mathcal{H}}(t) = \hat{\mathcal{H}}_0 + \hat{\mathcal{H}}_1(t)$, where $\hat{\mathcal{H}}_0$ is a time-independent part, and $\hat{\mathcal{H}}_1(t) = \hat{\mathcal{H}}_1(t+T)$ is periodic with a period $T = 2\pi/\omega$. The corresponding unitary operator for a single period can be rewritten as an effective evolution with a time-independent Hamiltonian given by the Magnus expansion \cite{Bukov2015,Iadecola2015}. When the frequency of modulation $\omega$ is much bigger than the coupling in the static Hamiltonian, $\omega / \Vert \hat{\mathcal{H}}_0 \Vert \gg 1$, the dynamics of the system can be conveniently represented in terms of a period averaged Floquet Hamiltonian,
\begin{equation}
\label{eq:H_F_Magnus}
\hat{\mathcal{H}}_{\mathrm{F}} = \frac{1}{T} \int_{0}^{T} \hat{\mathcal{H}}'_0(t) dt,
\end{equation}
where $\hat{\mathcal{H}}'_0(t)$ denotes the static Hamiltonian $\hat{\mathcal{H}}_0$ rewritten in the interaction picture with respect to the oscillating part. The resulting Hamiltonian $\hat{\mathcal{H}}_{F}$ may possess qualitatively different behavior compared to $\hat{\mathcal{H}}_0$. Such Floquet type simulation recently gained attention in the cold atom \cite{Goldman2014} and condensed matter physics communities \cite{Rudner2013,Jiang2011}, where Floquet topological insulators and gauge fields were introduced. 
Finally, in circuit QED the Floquet quasienergies of a single qubit were studied recently \cite{Deng2015}.

In this paper we consider superconducting qubits with isotropic XY coupling, and show that by exploiting fast driving of each site we can efficiently tailor the effective Hamiltonian of the system. For concreteness we here focus on one dimensional chains, but the method can easily be extended to more dimensions. First, we describe how the approach can be used to simulate the dynamics of the transverse Ising model showing that as compared to digital protocols higher fidelity can be attained. Next, we simulate quantum annealing to the ground state of the transverse Ising model and find that the Floquet approach outperforms the digital for restricted resources, when limited by the finite anharmonicity of the transmons. Finally, we describe an algorithm for simulating the spin-1/2 XYZ model with the Floquet approach.

\textit{System and Hamiltonian.---}As a physical realization we consider a capacitively coupled chain of qubits (Fig. \ref{fig:sketch}), where periodically oscillating effective magnetic fields are applied at chosen lattice sites. The qubits can be of transmon \cite{Riste2015}, xmon \cite{Barends2013}, gmon \cite{Chen2014}, or gatemon-type \cite{Casparis2016}, with the main requirements being tunability and high fidelity operation. The time independent Hamiltonian in the rotating frame $\hat{\mathcal{H}}_0$ contains a nearest-neighbour flip-flop interaction with bare coupling $J$, corresponding to the isotropic XY spin-1/2 model
\begin{equation}
\label{eq:Hamiltonian_0_and_1}
\hat{\mathcal{H}}_0 = \sum\limits_{j=1}^{N-1} J (\sigma^x_{j} \sigma^x_{j+1} + \sigma^y_{j} \sigma^y_{j+1}).
\end{equation}
Here $\sigma^\alpha_j$ ($\alpha = x,y,z$) are spin-1/2 Pauli operators at lattice site $j$, $N$ denotes total number of qubits in the chain, and we consider open boundary conditions. The time dependent Hamiltonian $\hat{\mathcal{H}}_1(t)$ contains a periodic magnetic field $\mathbf{h}_j(t)$ which rapidly oscillates along arbitrary axes, and we assume that it differs between even and odd sites,
\begin{equation}
\label{eq:H_magn}
\hat{\mathcal{H}}_1(t) = \sum\limits_{j=1}^{\lceil N/2 \rceil} \mathbf{h}_{\mathrm{odd}}(t) \cdot \boldsymbol{\sigma}_{2j-1} + \sum\limits_{j=1}^{\lfloor N/2 \rfloor} \mathbf{h}_{\mathrm{even}}(t) \cdot \boldsymbol{\sigma}_{2j},
\end{equation}
where $\lfloor x \rfloor$ and $\lceil x \rceil$ denote floor and ceiling functions, respectively. We assume that the magnetic field is sharply turned on at time $t = 0$, and thus explicitly account for the kick operator contribution, which is relevant in the current setting targeting realistic quantum simulation \cite{Goldman2014}. Going to the rotating frame with respect to $\hat{\mathcal{H}}_1$ and integrating over a period as in Eq. (\ref{eq:H_F_Magnus}) we get the reduced Floquet Hamiltonian in the general form [\onlinecite{SM}A]:
\begin{equation}
\label{eq:H_Floquet_gen}
\hat{\mathcal{H}}_{\mathrm{F}} = \sum\limits_{\alpha,\alpha' = x,y,z} \sum\limits_{j=1}^{\lfloor N/2 \rfloor} \overline{\xi_{\alpha \alpha'}} \sigma^\alpha_{2j} (\sigma^{\alpha'}_{2j-1} + \sigma^{\alpha'}_{2j+1}),
\end{equation}
where the time-averaged coefficients $\overline{\xi_{\alpha \alpha'}}$ are defined in the Supplemental Material \cite{SM}, and are controlled by the amplitude and alignment of the effective magnetic fields.
\begin{figure}[t]
\includegraphics[width=1.0\linewidth]{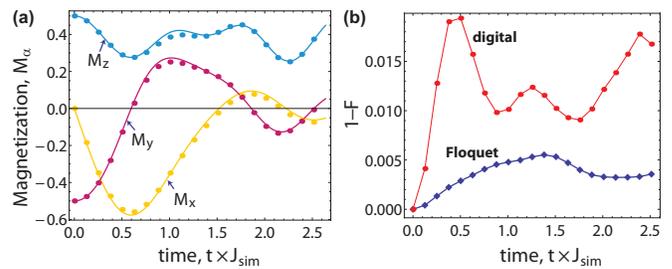}
\caption{\textbf{Transverse Ising dynamics.} (a) Total normalized magnetization of the $N=4$ chain $M_{\alpha}(t) = \langle \psi(t)| \sum_{j}^{N} \sigma^\alpha_j |\psi(t) \rangle / N$. Solid curves show the ideal continuous evolution under the transverse Ising Hamiltonian. Bullets correspond to stroboscopic periods of the Floquet dynamics, showing good agreement with the original model. We set $J<0$, $h^z/J = 3/2$, $\omega/|J| = 50$, and run the simulation for a total time of 20 stroboscopic periods. (b) Simulation infidelities. Blue diamonds show the dynamical overlap of the Floquet evolution with the ideal transverse Ising evolution, $F = |\langle \psi_{\mathrm{tIsing}}(nT)|\psi(nT)\rangle |^2$. The red bullets correspond to digital evolution with $N_{\mathrm{Tr}} = 20$ Trotter steps.}
\label{fig:tIsing-dyn}
\end{figure}

\textit{Transverse Ising model: quantum dynamics.---}We first consider the Floquet simulation of the tranverse Ising model, which represents a particular case of the more general Floquet Hamiltonian (\ref{eq:H_Floquet_gen}). It can be realized with a drive of the form $\mathbf{h}_{\mathrm{even}}(t) = \lambda \omega \cos(\omega t) \mathbf{e}^x + 2 h^z (1+\mathcal{J}_0[4 \lambda])^{-1} \cos(2 \lambda \sin[\omega t]) \mathbf{e}^z$ and $\mathbf{h}_{\mathrm{odd}}(t) = h^z \mathbf{e}^z$, where $\lambda$ is a drive parameter, and $\mathcal{J}_0[x]$ denotes the zeroth order Bessel function of the first kind. The $z$-directed terms additionally introduce an effective transverse magnetic field $h^z$. In the infinite frequency limit $|J|/\omega \rightarrow 0$ and for $\lambda = 1.20241$ ($\mathcal{J}_0[2 \lambda] = 0$) this leads to an effective Hamiltonian of the form [\onlinecite{SM}B]
\begin{align}
\label{eq:H_tIsing}
\hat{\mathcal{H}}_{F} = J_{\mathrm{sim}} \sum\limits_{j=1}^{N-1} \sigma_j^x \sigma_{j+1}^x + h^z \sum\limits_{j=1}^{N} \sigma_j^z \equiv \hat{\mathcal{H}}_{\mathrm{tIsing}},
\end{align}
where for later reference $J_{\mathrm{sim}}$ describes the effective simulated coupling of the model. In the transverse Ising case it corresponds to the original bare coupling, $J_{\mathrm{sim}} = J$, and we note that the ratio between the effective magnetic field in the Floquet basis and the Ising interaction, $h^z/J_{\mathrm{sim}}$, can be controlled by the drive parameters.
To verify the approach, we numerically calculate the full dynamics of the system with time periodic driving of frequency $\omega / |J| = 50$ and access the Floquet dynamics by looking at stroboscopic times $n T$, where $n$ is an integer [Fig. \ref{fig:tIsing-dyn} (a)]. The results are compared to an ideal simulation of the transverse Ising Hamiltonian (\ref{eq:H_tIsing}), with the initial state $|\psi_{\mathrm{in}}\rangle = \bigotimes_{j=1}^{N/2} (|\uparrow\rangle_{2j-1} - i |\downarrow\rangle_{2j-1})/\sqrt{2} \otimes |\uparrow\rangle_{2j}$. 
Additionally, we benchmark the Floquet simulation with a digital protocol [\onlinecite{SM}C]. It uses the isotropic XY interaction and its rotated version, and approaches the Ising model in the limit of large $N_{\mathrm{Tr}}$ \cite{LasHeras2014}. The results in Fig. \ref{fig:tIsing-dyn}(b) show that the Floquet simulation closely follows the exact dynamics at short time, but deviates at later stages due to the finite Floquet frequency. On the contrary, the digital approach shows substantial deviations for this limited number of Trotter periods even at short times, but will have rapid convergence with more Trotter steps (see below). We note that a comparison of the number of Floques periods and Trotter steps may not be a fair comparison, since the latter involves multiple gates. A more detailed comparison is done below.
\begin{figure}[t]
\includegraphics[width=1.0\linewidth]{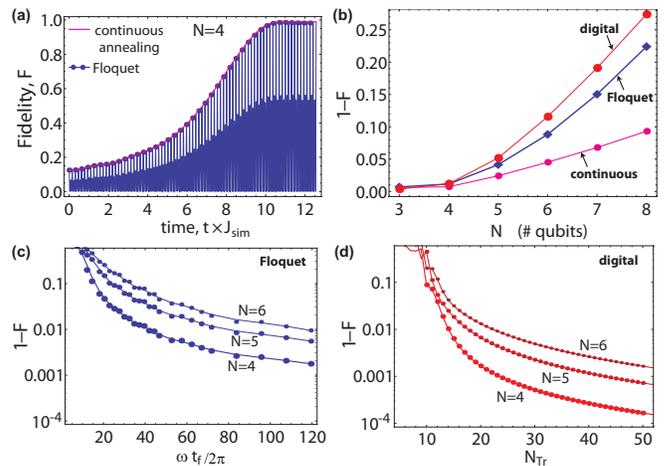}
\caption{\textbf{Transverse Ising annealing.} (a) Fidelity of the simulated state with respect to the ideal target state. The dark red line denotes annealing with the ideal transverse Ising Hamiltonian. The blue thin line corresponds to evolution under the time-dependent Hamiltonian, with blue dots showing state fidelities at accessible stroboscopic times. We assume $J_{\mathrm{sim}}<0$, $h^z/J_{\mathrm{sim}} = 1$, $t_f = 12.5 |J_{\mathrm{sim}}|^{-1}$, and $\omega/|J| = 20$. (b) Final time infidelities for ideal continuous, Floquet, and digital evolution shown for different number of qubits. Parameters are same as in (a) with final simulation time $t_f = 15 |J_{\mathrm{sim}}|^{-1}$. The digital evolution corresponds to $N_{\mathrm{Tr}} = 20$ Trotter steps. (c) Final fidelity of Floquet simulation, measured with respect to finite time annealing, and plotted as a function of modulation frequency for $N=4,5,6$. (d) Fidelity of the digital annealing with respect to the continuous annealing state fidelity plotted as function of the number of Trotter steps for $N=4,5,6$ qubits.}
\label{fig:tIsing-annl}
\end{figure}

\textit{Transverse Ising model: quantum annealing.---}Next, we study the ground state preparation of the simulated model. To access a ground state, we perform the quantum annealing procedure \cite{Kadowaki1998}, which also serves as a basis for adiabatic quantum computing \cite{Fahri2000}, and may solve NP-complete problems \cite{Fahri2001}. Using the Floquet basis, we design the Hamiltonian $\hat{\mathcal{H}}_{F}(t) = J_{\mathrm{sim}} \sum_{j=1}^{N-1} \sigma_j^x \sigma_{j+1}^x + (1 - t/t_f) h^z \sum_{j=1}^{N} \sigma_j^z$, where the effective magnetic field is linearly turned off during a total annealing time $t_f$. Here we consider $J_{\mathrm{sim}}<0$, which allows to achieve the ground state of the ferromagnetic $x$-Ising Hamiltonian. The ideal target state is an entangled GHZ-like state $|\psi_\mathrm{T} \rangle = (|+\rangle^{\otimes N} + |-\rangle^{\otimes N})/\sqrt{2}$, where $|\pm\rangle = {(|\downarrow\rangle \pm |\uparrow\rangle)}/\sqrt{2}$, and we start from the trivial initial state $|\psi_\mathrm{in} \rangle = |\downarrow\rangle^{\otimes N}$.

The results of the annealing procedure are shown in Fig. \ref{fig:tIsing-annl}. The dynamics of the system, quantified as the fidelity of the instantaneous wavefunction of the system with the ideal target state,  $F = |\langle \psi_{\mathrm{T}}|\psi(t)\rangle |^2$, is shown in Fig. \ref{fig:tIsing-annl}(a) for a 4 qubits chain. Blue dots correspond to Floquet simulation at stroboscopic times, which closely follows the red solid curve of the ideal continuous annealer. The blue oscillatory curve corresponds to the full dynamics, with fast oscillation arising from the drive term.
To study the scaling with the system size, we perform fixed time ($t_f = 15 |J_{\mathrm{sim}}|^{-1}$) annealing for chains of various length [Fig. \ref{fig:tIsing-annl}(b)]. Fixing the drive frequency to moderate values, we observe that the final infidelities of the Floquet simulator, $1 - F_{\mathrm{Floquet}}(t_f)$, and digital simulator, $1 - F_{\mathrm{digital}}(t_f)$, have similar scalings with the system size, both adding extra infidelity on top of the continuous evolution, and largely dependent on $\omega$ and $N_{\mathrm{Tr}}$ as described below.
In Fig. \ref{fig:tIsing-annl}(c) we show the dependence on the frequency of the periodic drive of the Floquet infidelity, measured with respect to the finite time annealing state. Here the frequency is rescaled by the total annealing time, such that $\omega t_f/2\pi$ shows the number of stroboscopic periods. The infidelity can be reduced for large $\omega$, with results converging to continuous annealing infidelity for $\omega/|J| \rightarrow \infty$. The analogous behavior for the digital approach corresponds to the variation of the number of Trotter steps, and is shown in Fig. \ref{fig:tIsing-annl}(d). While a direct comparison between two approaches is complicated, the general tendency can be deduced: the Floquet approach has smaller infidelity for small number of steps and limited resources, while the digital approach has better scaling if a large number of Trotter steps $N_{\mathrm{Tr}}$ can be implemented.
\begin{figure}[t]
\includegraphics[width=1.0\linewidth]{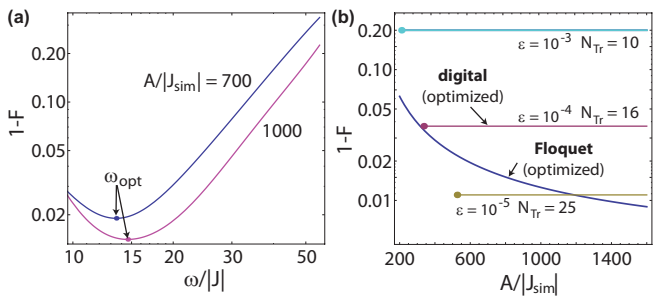}
\caption{\textbf{Imperfections.} (a) Infidelity of Floquet transverse Ising annealing ($N=4$), measured with respect to the ideal target state, and calculated for fixed anharmonicity and varying frequency. The point of low infidelity defines an optimal frequency for the simulation. (b) Final state infidelity for transverse Ising annealing for $N = 4$, $\omega = \omega^{\mathrm{opt}}$, and $t_f = 15 |J|^{-1}$. Here the anharmonicity $A$ spans the range $A/\omega = {19-99}$. Horizontal lines show optimized digital protocol infidelities for fixed single gate error $\epsilon$, with cut-off (large dots) determined by a minimal gate time $t_{\mathrm{gate}} \geq  35A^{-1}$.}
\label{fig:leakage}
\end{figure}

\textit{Imperfections.---}To describe a realistic quantum simulator, we study the influence of a finite anharmonicity $A$ of the transmon qubits, which will be a major limitation to our approach. Driving a transmon with a finite anharmonicity leads to leakage of information from the logical subspace. To account for this we consider a full Hamiltonian of a SC chain [\onlinecite{SM}D], and perform numerical simulations including doubly occupied states of the transmons. As an example we use annealing of the transverse Ising model with $N=4$. The resulting infidelity of the simulation is shown in Fig. \ref{fig:leakage}. First, we fix the value of the anharmonicity, and calculate the infidelity as a function of the Floquet frequency [Fig. \ref{fig:leakage}(a)]. We observe that contrary to the ideal circuit, the infidelity is minimized for a finite (optimal) drive frequency $\omega^{\mathrm{opt}}$. We note that the window of frequencies in which the infidelity stays close to minimal is typically broad. In Fig. \ref{fig:leakage}(b) we show the optimized infidelity of the simulation as a function of $A$ (blue curve).

To benchmark the results of the Floquet simulation we compare it to the digital simulation. Assuming the same value $\epsilon$ for the single- and two-qubit gate errors, the error for a single Trotter step is equal to $\epsilon_{\mathrm{Tr}} = (5 N - 4) \epsilon$. We optimize the total infidelity $1-F_{\mathrm{tot}} = 1 - (1 - \epsilon_{\mathrm{Tr}})^{N_{\mathrm{Tr}}} (1 - \epsilon_{\mathrm{dig}})$ with respect to the number of Trotter steps. Here $\epsilon_{\mathrm{dig}}$ is the error coming from digitization of the dynamics [as shown in Fig. \ref{fig:tIsing-annl}(d)]. To highlight the relevant parameter range, we assign a cut-off based on the single gate time $t_{\mathrm{gate}} \geq 35 A^{-1}$. The results are plotted as horizontal lines in Fig. \ref{fig:leakage}(b), and show that the Floquet approach can outperform the  digital approach unless very high fidelity gates with $\epsilon < 10^{-4}$ are used. Furthermore, the Floquet approach is highly advantageous for small values of $A/J_{\mathrm{sim}}$, which for a given transmon anharmonicity $A$ is the regime where the simulation is finished the fastest and thus has the least influence of decoherence. 
Thus for shorter time of simulation/higher error rates the Floquet is advantageous compared to the digital approach. We highlight that while digital approaches typically exploit DRAG techniques to remove leakage \cite{Motzoi2009,Martinis2014}, the presented Floquet approach is not specifically designed to work for small $A$, and it may be possible to improve on this issue using few-tone drives.

\textit{Estimates.---}To quantify the performance we consider numbers which can be achieved with currently available transmon setups \cite{Barends2013,Chen2014}. Taking the anharmonicity to be $A = 2\pi \times 300$ MHz, drive frequency $\omega = 2\pi \times 9.8$ MHz, nearest-neighbour coupling $J = 2 \pi \times 1$ MHz (reduced compared to most setups), and $t_f = 2.4~\mu$s, the four qubit chain can be annealed to the ground state of the Ising model with $1-F_{\mathrm{Flq}} = 0.037$ (ideal continuous annealing gives $1-F_{\mathrm{cont}} = 0.00616$). An additional error will arise from dephasing, but its effect will be small for highly coherent qubits ($T_2 > 10~\mu$s).

Reaching a similar performance with the digital strategy is highly challenging and would require single and two qubit gate operation times of $18$~ns~$=33 A^{-1}$ and $\epsilon = 10^{-4}$ accuracy. The single Trotter step duration for $N=4$ is $t_{\mathrm{Tr}} = 0.162~\mu$s, and with $t_f = 2.4~\mu$s this can allow for 14 Trotter steps. The corresponding digitization error is $1 - F_{\mathrm{dig}} = 0.041$ (dephasing should be added separately).

\textit{Generic XYZ Hamiltonians.---}The Floquet approach may be extended to simulate generic spin-1/2 models represented by XYZ type spin Hamiltonian, $\hat{\mathcal{H}}_{\mathrm{XYZ}} = \sum_{j=1}^{N-1} (J^x \sigma_j^x \sigma_{j+1}^x + J^y \sigma_j^y \sigma_{j+1}^y + J^z \sigma_j^z \sigma_{j+1}^z)$. These are of the so called non-stoquatic type, where recent results have suggested that they can give enhanced computational powers \cite{Nishimori2016}. We consider couplings $J_y = 2 J^x/ 3$, $J^z = J^x/3$, and $J^x = J < 0$. This configuration can be realized by a uniform periodic magnetic field in the $x$ direction at all lattice sites [\onlinecite{SM}E]. As compared to the transverse Ising case this Hamiltonian possesses small energy gaps, and in the absence of the additional transverse field is difficult to anneal even with the ideal continuous Hamiltonian. The results are qualitatively similar to the Ising case, showing that the Floquet approach outperforms digital simulation unless ultrahigh fidelity gates are implemented, although high-quality simulation is very challenging in both cases.

\textit{Conclusion.---}We have presented a scheme for a reconfigurable and tunable superconducting quantum simulator based on the transmon qubits. Utilizing the Floquet approach, we show that the originally limited isotropic XY interaction can be transformed into transverse Ising or XYZ type spin-1/2 Hamiltonian. The approach was used for simulation of multiqubit system dynamics and preparation of non-trivial groundstates. The Floquet simulation was shown to perform better than a digital scheme for restricted resources, and represents a realistic path for modern SC quantum simulators.

\textit{Acknowledgements.---}The research was funded by the European Union Seventh Framework Programme through ERC Grant QIOS (Grant No. 306576). O.K. acknowledges useful discussions with Pedram Roushan, Charles Neill, Guanyu Zhu, Mohammad Hafezi, G\"{o}ran Johansson, and Stefan Filipp.

\onecolumngrid
\newpage

\setcounter{equation}{0}
\setcounter{figure}{0}
\renewcommand{\theequation}{S\arabic{equation}}
\renewcommand{\thefigure}{S\arabic{figure}}

\begin{center}
\textbf{Supplemental Material: Floquet quantum simulation with superconducting qubit chains}
\end{center}

\tableofcontents

\subsection{\label{sect:A}Derivation of the generic Floquet Hamiltonian}

Here we present the derivation of Eq. (4) in the main text, and simultaneously describe the general Floquet Hamiltonian originating from arbitrary axis magnetic field oscillation for the two sublattices.

We start with the transmon Hamiltonian written in the form
\begin{align}
\label{eq:H_start}
\hat{\mathcal{H}}(t) = \hat{\mathcal{H}}_0 + \hat{\mathcal{H}}_1(t) = J \sum\limits_{j=1}^{N-1}(\sigma^x_{j} \sigma^x_{j+1} + \sigma^y_{j} \sigma^y_{j+1}) + \sum\limits_{j=1}^{\lceil N/2 \rceil} \mathbf{h}_{\mathrm{odd}}(t) \cdot \boldsymbol{\sigma}_{2j-1} + \sum\limits_{j=1}^{\lfloor N/2 \rfloor} \mathbf{h}_{\mathrm{even}}(t) \cdot \boldsymbol{\sigma}_{2j},
\end{align}
and consider the effective magnetic fields to be different for even and odd sublattices. Here $\lfloor x \rfloor$ and $\lceil x \rceil$ denote floor and ceiling functions, correspondingly.
To have an explicit form of $\mathbf{h}_{\mathrm{even/odd}}(t)$ we decompose it into $\mathbf{h}_{\mathrm{even/odd}}(t) = f_{\mathrm{e/o}}(t) (\mathbf{e}^x h^x_{\mathrm{e/o}} + \mathbf{e}^y h^y_{\mathrm{e/o}} + \mathbf{e}^z h^z_{\mathrm{e/o}} )$, where $f_{\mathrm{e/o}}(t) = f_{\mathrm{e/o}}(t+T)$ is some periodic function, $\mathbf{e}^{x,y,z}$ form a Cartesian basis, and $h^{\alpha}_{\mathrm{e/o}}$ are components of an effective magnetic field for even (indexed by $_\mathrm{e}$) and odd (indexed by $_{\mathrm{o}}$) sites. This corresponds to the magnetic field oscillating along a certain axis, which is different for even and odd sublattices. Here we consider the time-dependence to be the same for all spin components, such that the axis of magnetic field does not precess. This restriction decreases the number of independent tuning parameters. Note that the effective magnetic field used for the simulation of the transverse Ising model, deviates slightly from this form. This deviation is, however, a perturbation and will be dealt with below. If the fields do not have this form, the time-dependent Hamiltonian $\hat{\mathcal{H}}_{1}(t)$ does not commute with itself at different times, thus largely complicating the solution. This more general setting will be considered in future works.

We perform a unitary transformation with respect to the rapidly oscillating time-dependent part $\hat{\mathcal{H}}_1(t)$. This is done using the unitary operator:
\begin{equation}
\label{eqS:U_t}
\hat{\mathcal{U}}(t) = \mathcal{T} \exp\left[-i \int\limits_{t_0}^{t} dt' f_{\mathrm{o}}(t') \sum\limits_{j=1}^{\lceil N/2 \rceil} (h^x_{\mathrm{o}} \sigma^x_{2j-1} + h^y_{\mathrm{o}} \sigma^y_{2j-1} + h^z_{\mathrm{o}} \sigma^z_{2j-1}) + f_{\mathrm{e}}(t') \sum\limits_{j=1}^{\lfloor N/2 \rfloor} (h^x_{\mathrm{e}} \sigma^x_{2j} + h^y_{\mathrm{e}} \sigma^y_{2j} + h^z_{\mathrm{e}} \sigma^z_{2j})\right],
\end{equation}
where $\hat{\mathcal{T}}\{...\}$ is the time-ordering operator and $t_0$ is an initial switch-on time. The unitary (\ref{eqS:U_t}) can be factorized into odd and even sublattices parts, and time-ordering disappears as the magnetic field oscillates along a fixed axis. This yields $\hat{\mathcal{U}}(t) = \hat{\mathcal{U}}_{\mathrm{o}}(t) \hat{\mathcal{U}}_{\mathrm{e}}(t)$ with
\begin{equation}
\hat{\mathcal{U}}_{\mathrm{p}}(t) = \exp \left[ -i \frac{g_{\mathrm{p}}(t)}{2} \sum\limits_{j=1}^{N_{\mathrm{p}}} (n^x_{\mathrm{p}} \sigma^x_{P} + n^y_{\mathrm{p}} \sigma^y_{P} + n^z_{\mathrm{p}} \sigma^z_{P}) \right],
\end{equation}
where indices $\mathrm{p} = \mathrm{o,e}$ and $\mathrm{P} = \mathrm{2j-1,2j}$ help denote sublattice parity, and we define the time-integrated functions $g_{\mathrm{p}}(t) \equiv 2 \int\limits_{t_0}^{t} dt' f_{\mathrm{p}}(t') h_{\mathrm{p}}$. Here $h_{\mathrm{p}} = \sqrt{(h^x_{\mathrm{p}})^2 + (h^y_{\mathrm{p}})^2 + (h^z_{\mathrm{p}})^2}$ defines the absolute value of the effective magnetic field vector, and $n^{x,y,z}_{\mathrm{p}} = h^{x,y,z}_{\mathrm{p}}/h_{\mathrm{p}}$ correspond to the normalized Cartesian components. The rotated Hamiltonian then reads $\hat{\mathcal{H}}'(t) = \hat{\mathcal{U}}^{\dagger}(t) \hat{\mathcal{H}}(t) \hat{\mathcal{U}}(t) - i \hat{\mathcal{U}}^{\dagger}(t)\partial_t \hat{\mathcal{U}}(t)$. The second term is divided into two parts and each of these can be rewritten using the relation for the derivative of an arbitrary time-dependent matrix $A(t)$, being $e^{-A(t)}\partial_t e^{A(t)} = \dot{A}(t) - [A(t), \dot{A}(t)]/2! + [A(t), [A(t), \dot{A}(t)]]/3! - ...$. For the case of $A(t) = -i \frac{g_{\mathrm{p}}(t)}{2} \sum\limits_{j=1}^{N_{\mathrm{p}}} (n^x_{\mathrm{p}} \sigma^x_{P} + n^y_{\mathrm{p}} \sigma^y_{P} + n^z_{\mathrm{p}} \sigma^z_{P})$ considered here the commutators vanish, and in total the derivative term gives $-\hat{\mathcal{H}}_1(t)$. However, we emphasize that this conclusion would not be true for general time-dependence of the Cartesian components of an effective magnetic field, where additional derivative-dependent terms appear.

Next, we need to calculate the matrix product terms of the form
\begin{align}
& \hat{\mathcal{U}}^{\dagger}(t) \left[ J \sum\limits_{j=1}^{N-1}(\sigma^x_{j} \sigma^x_{j+1} + \sigma^y_{j} \sigma^y_{j+1}) \right] \hat{\mathcal{U}}(t) = e^{\left[ i \frac{g_{\mathrm{e}}(t)}{2} \sum\limits_{j=1}^{N_{\mathrm{e}}} (n^x_{\mathrm{e}} \sigma^x_{2j} + n^y_{\mathrm{e}} \sigma^y_{2j} + n^z_{\mathrm{e}} \sigma^z_{2j})\right] } e^{\left[ i \frac{g_{\mathrm{o}}(t)}{2} \sum\limits_{j=1}^{N_{\mathrm{o}}} (n^x_{\mathrm{o}} \sigma^x_{2j-1} + n^y_{\mathrm{o}} \sigma^y_{2j-1} + n^z_{\mathrm{o}} \sigma^z_{2j-1}) \right]} \times \\ \notag & \times \left[ J \sum\limits_{j=1}^{N-1}(\sigma^x_{j} \sigma^x_{j+1} + \sigma^y_{j} \sigma^y_{j+1}) \right] e^{\left[ -i \frac{g_{\mathrm{o}}(t)}{2} \sum\limits_{j=1}^{N_{\mathrm{o}}} (n^x_{\mathrm{o}} \sigma^x_{2j-1} + n^y_{\mathrm{o}} \sigma^y_{2j-1} + n^z_{\mathrm{o}} \sigma^z_{2j-1}) \right]} e^{\left[-i \frac{g_{\mathrm{e}}(t)}{2} \sum\limits_{j=1}^{N_{\mathrm{e}}} (n^x_{\mathrm{e}} \sigma^x_{2j} + n^y_{\mathrm{e}} \sigma^y_{2j} + n^z_{\mathrm{e}} \sigma^z_{2j}) \right]}.
\end{align}
Let us perform the unitary rotation for each sublattice consecutively. For this, it is convenient to go from Cartesian to the spherical coordinate frame $\mathbf{n}_{\mathrm{p}} = \{n^x_{\mathrm{p}}, n^y_{\mathrm{p}}, n^z_{\mathrm{p}} \} \leftrightarrow \{ \cos\phi_{\mathrm{p}} \sin\theta_{\mathrm{p}}, \sin\phi_{\mathrm{p}} \sin\theta_{\mathrm{p}}, \cos\theta_{\mathrm{p}} \}$. Next, noting that the unitary operation is defined as a spin rotation with respect to a fixed axis $\mathbf{n}_{\mathrm{p}}$, it can be decomposed into Cartesian rotations as:
\begin{equation}
\label{eqS:Rn}
\hat{\mathcal{R}}^{\mathbf{n}}_{j}(g) = e^{-i \frac{g}{2} \mathbf{n}\cdot \sigma_j} = \hat{\mathcal{R}}^{z}_{j}(\phi) \hat{\mathcal{R}}^{y}_{j}(\theta) \hat{\mathcal{R}}^{z}_{j}(g) \hat{\mathcal{R}}^{y}_{j}(\theta)^\dagger \hat{\mathcal{R}}^{z}_{j}(\phi)^\dagger,
\end{equation}
where the rotation operator is defined as $\hat{\mathcal{R}}^{\alpha}(\varphi) = e^{-i \frac{\varphi}{2} \sigma^{\alpha}}$ ($\alpha = x,y,z$). Then, the Hamiltonian after the odd sublattice transformation can be obtained through rotations
\begin{equation}
\label{eqS:rotation_series}
\hat{\mathcal{U}}^{\dagger}(t) \hat{\mathcal{H}}_0 \hat{\mathcal{U}}(t) = \hat{\mathcal{U}}_{\mathrm{e}}^{\dagger} \big[ \hat{\mathcal{R}}^{z}_{\mathrm{o}}(\phi_{\mathrm{o}}) \hat{\mathcal{R}}^{y}_{\mathrm{o}}(\theta_{\mathrm{o}}) \hat{\mathcal{R}}^{z}_{\mathrm{o}}(-g_{\mathrm{o}}) \hat{\mathcal{R}}^{y}_{\mathrm{o}}(-\theta_{\mathrm{o}}) \hat{\mathcal{R}}^{z}_{\mathrm{o}}(-\phi_{\mathrm{o}}) \hat{\mathcal{H}}_0 \hat{\mathcal{R}}^{z}_{\mathrm{o}}(\phi_{\mathrm{o}}) \hat{\mathcal{R}}^{y}_{\mathrm{o}}(\theta_{\mathrm{o}}) \hat{\mathcal{R}}^{z}_{\mathrm{o}}(g_{\mathrm{o}}) \hat{\mathcal{R}}^{y}_{\mathrm{o}}(-\theta_{\mathrm{o}}) \hat{\mathcal{R}}^{z}_{\mathrm{o}}(-\phi_{\mathrm{o}}) \big] \hat{\mathcal{U}}_{\mathrm{e}},
\end{equation}
and the subsequent transformation for the even sublattice can be performed in a similar fashion.

Finally, to get a closed expression for the transformed Hamiltonian we use the Baker-Campbell-Hausdorff formula
\begin{equation}
\label{eqS:BCH}
e^{\hat{\mathcal{M}}} \hat{\mathcal{H}}_0 e^{-\hat{\mathcal{M}}} = \hat{\mathcal{H}}_0 + [\hat{\mathcal{M}}, \hat{\mathcal{H}}_0] + \frac{1}{2!}[\hat{\mathcal{M}}, [\hat{\mathcal{M}}, \hat{\mathcal{H}}_0]] + ... = \sum\limits_{k=0}^{\infty} \frac{1}{k!} [\hat{\mathcal{M}}, \hat{\mathcal{H}}_0]_k,
\end{equation}
where $[\hat{\mathcal{M}}, \hat{\mathcal{H}}_0]_k$ denotes $k$-th order nested commutator. We proceed with calculating the commutators and resumming the series. After straightforward but tedious algebra we can get the Hamiltonian in a rotating frame
\begin{equation}
\label{eqS:Ht_rot}
\hat{\mathcal{H}}'(t) = \sum\limits_{\alpha,\alpha' = x,y,z} \sum\limits_{j=1}^{N/2} \xi_{\alpha \alpha'} \sigma^\alpha_{2j} (\sigma^{\alpha'}_{2j-1} + \sigma^{\alpha'}_{2j+1}),
\end{equation}
with the coefficients
\begin{align*}
&\xi_{xx}(t) = \sin^2(\theta_e) \cos(\phi_e) \Big[\cos(g_o[t]) \cos^2(\theta_o) \cos(\phi_o) \cos(\phi_e-\phi_o)-\sin(\phi_e) \sin(g_o[t]) \cos(\theta_o)-\cos(g_o[t]) \sin(\phi_o) \sin(\phi_e-\phi_o) \\ 
&+\sin^2(\theta_o) \cos(\phi_o) \cos(\phi_e-\phi_o)\Big] + \cos(g_e[t]) \Big\{\cos(g_o[t]) \Big[\cos^2(\theta_o) \cos(\phi_o) \big(\cos^2(\theta_e) \cos(\phi_e) \cos(\phi_e-\phi_o) + \sin(\phi_e) \times \\ \times & \sin(\phi_e-\phi_o)\big) + \sin(\phi_o) \big(\sin(\phi_e) \cos(\phi_e-\phi_o)-\cos^2(\theta_e) \cos(\phi_e) \sin(\phi_e-\phi_o)\big)\Big] + \sin^2(\theta_e) \sin(\phi_e) \cos(\phi_e) \sin(g_o[t]) \cos(\theta_o) \\ & + \cos^2(\theta_e) \cos(\phi_e) \sin^2(\theta_o) \cos(\phi_o) \cos(\phi_e-\phi_o) + \sin(\phi_e) \sin^2(\theta_o) \cos(\phi_o) \sin(\phi_e-\phi_o)\Big\} + \sin(g_e[t]) \cos(\theta_e) \times \\ & \times \Big[\sin(g_o[t]) \cos(\theta_o) -\sin^2\left(\frac{g_o[t]}{2}\right) \sin^2(\theta_o) \sin(2 \phi_o)\Big],
\end{align*}
\begin{align*}
&\xi_{yy}(t) = \sin^2(\theta_e) \sin(\phi_e) \Big[\sin(\phi_o) \cos(\phi_e-\phi_o) \big(\cos(g_o[t]) \cos^2(\theta_o)+\sin^2(\theta_o)\big) + \cos(\phi_e) \sin(g_o[t]) \cos(\theta_o) \\ &+ \cos(g_o[t]) \cos(\phi_o) \sin(\phi_e-\phi_o)\Big] + \cos(g_e[t]) \Big\{\cos(g_o[t]) \Big[\cos^2(\phi_e) \big(\cos^2(\theta_o) \sin^2(\phi_o) + \cos^2(\phi_o)\big) + \cos^2(\theta_e) \sin(\phi_e) \times \\ & \times \big(\cos^2(\theta_o) \sin(\phi_o) \cos(\phi_e-\phi_o) + \cos(\phi_o) \sin(\phi_e-\phi_o)\big) + \sin(\phi_e) \cos(\phi_e) \sin^2(\theta_o) \sin(\phi_o) \cos(\phi_o)\Big] - \sin^2(\theta_e) \sin(\phi_e) \times \\ & \times \cos(\phi_e) \sin(g_o[t]) \cos(\theta_o) + \sin^2(\theta_o) \sin(\phi_o) \big(\cos^2(\theta_e) \sin(\phi_e) \cos(\phi_e-\phi_o) - \cos(\phi_e) \sin(\phi_e-\phi_o)\big)\Big\} + \sin(g_e[t]) \cos(\theta_e) \times \\ & \times \Big[\sin(g_o[t]) \cos(\theta_o) + \sin^2\left(\frac{g_o[t]}{2}\right) \sin^2(\theta_o) \sin(2\phi_o) \Big],
\end{align*}
\begin{align*}
&\xi_{zz}(t) = \sin(\theta_e) \sin(\theta_o) \bigg\{ \cos(\phi_e-\phi_o) \left[\sin(g_e[t]) \sin(g_o[t]) + 2 \big(1 - \cos(g_e[t])\big) \cos(\theta_e) \sin^2\left(\frac{g_o[t]}{2}\right) \cos(\theta_o)\right] \\ &+ \sin(\phi_e-\phi_o) \left[ \big(1 - \cos(g_e[t])\big) \cos(\theta_e) \sin(g_o[t]) - 2 \sin(g_e[t]) \sin^2\left(\frac{g_o[t]}{2}\right) \cos(\theta_o) \right] \bigg\},
\end{align*}
\begin{align*}
&\xi_{xy}(t) = \sin^2(\theta_o) \sin(\phi_o) \Big[\cos(g_e[t]) \Big(\cos^2(\theta_e) \cos(\phi_e) \cos (\phi_e-\phi_o) + \sin(\phi_e) \sin(\phi_e-\phi_o)\Big) + \sin^2(\theta_e) \cos^2(\phi_e) \cos(\phi_o)\Big] \\ &+ \sin^2(\phi_o) \Big[\sin^2(\theta_e) \cos^2(\phi_e) \sin(g_o[t]) \cos(\theta_o) + \frac{1}{2} \sin^2(\theta_o) \Big(\sin^2(\theta_e) \sin(2\phi_e) - 2 \sin(g_e[t]) \cos(\theta_e)\Big)\Big] + \sin(g_o[t]) \cos(\theta_o) \times \\ & \times \Big[\cos^2(\phi_e) \Big(\cos(g_e[t]) \cos^2(\theta_e) + \sin^2(\theta_e) \cos^2(\phi_o)\Big) + \cos(g_e[t]) \sin^2(\phi_e)\Big] + \frac{1}{4} \cos(g_o[t]) \Bigg[4 \cos(g_e[t]) \cos^2(\theta_e) \cos(\phi_e) \cos(\phi_o) \times \\ & \times \Big(\sin(\phi_e) \cos(\phi_o)-\cos(\phi_e) \sin^2(\theta_o) \sin(\phi_o)\Big) + 4 \sin^2\left(\frac{g_e[t]}{2}\right) \sin^2(\theta_e) \sin(2 \phi_e) \cos^2(\theta_o) \sin^2(\phi_o) + 4 \sin(\phi_e) \cos(\phi_e) \times \\ & \times \cos^2(\phi_o) \Big(\sin^2(\theta_e) -\cos(g_e[t])\Big) - 4 \sin^2(\theta_o) \sin(\phi_o) \cos(\phi_o) \Big(\cos(g_e[t]) \sin^2(\phi_e)+\sin^2(\theta_e) \cos^2(\phi_e)\Big) - \sin(g_e[t]) \cos(\theta_e) \times \\ & \times \Big(2 \sin^2(\theta_o) \cos(2 \phi_o) + \cos(2 \theta_o)+3\Big)\Bigg],
\end{align*}
\begin{align*}
&\xi_{yx}(t) = \frac{1}{8} \Bigg\{8 \sin(g_e[t]) \cos(\theta_e) \sin^2(\theta_o) \cos^2(\phi_o) + 8 \cos(g_e[t]) \sin^2(\theta_o) \sin(\phi_o) \cos(\phi_o) \Big(\cos^2(\theta_e) \sin^2(\phi_e) + \cos^2(\phi_e)\Big) \\ &+ 2 \cos(g_o[t]) \bigg[4 \cos^2(\theta_o) \cos(\phi_o) \Big(\sin^2(\theta_e) \sin(\phi_e) \cos(\phi_e-\phi_o) -\cos(g_e[t]) \cos(\phi_e) \sin(\phi_e-\phi_o)\Big) +4 \cos(g_e[t]) \cos^2(\theta_e) \times \\ &\times \sin(\phi_e) \Big(\cos^2(\theta_o) \cos(\phi_o) \cos(\phi_e-\phi_o) - \sin(\phi_o) \sin(\phi_e-\phi_o)\Big) + \sin(g_e[t]) \cos(\theta_e) \Big(-2 \sin^2(\theta_o) \cos(2\phi_o) +\cos(2\theta_o) + 3\Big) \\ &- 4 \sin(\phi_o) \Big(\cos(g_e[t]) \cos(\phi_e) \cos(\phi_e-\phi_o) +\sin^2(\theta_e) \sin(\phi_e) \sin(\phi_e-\phi_o)\Big) \bigg] + 8 \sin^2(\theta_e) \sin(\phi_e) \sin^2(\theta_o) \cos(\phi_o) \times \\ &\times \Big(\cos(\phi_e-\phi_o) -\cos(g_e[t]) \cos(\phi_e) \cos(\phi_o)\Big) -8 \sin^2(\theta_e) \sin^2(\phi_e) \sin(g_o[t]) \cos(\theta_o) - 2\cos(g_e[t]) \sin(g_o[t]) \cos(\theta_o) \times \\ &\times \Big(2 \sin^2(\theta_e) \cos(2\phi_e) + \cos(2\theta_e) + 3\Big) \Bigg\},
\end{align*}
\begin{align*}
&\xi_{yz}(t) = \sin(\theta_o) \Bigg\{\cos(g_e[t]) \cos^2(\theta_e) \sin(\phi_e) \Big(2\sin^2\left(\frac{g_o[t]}{2}\right) \cos(\theta_o) \cos(\phi_e-\phi_o) +\sin(g_o[t]) \sin(\phi_e-\phi_o)\Big) \\ &+\sin(g_e[t]) \cos(\theta_e) \Big(2\sin^2\left(\frac{g_o[t]}{2}\right) \cos(\theta_o) \cos(\phi_o)-\sin(g_o[t]) \sin(\phi_o)\Big) +\cos(\phi_e-\phi_o) \Big(2 \sin^2(\theta_e) \sin(\phi_e) \sin^2\left(\frac{g_o[t]}{2}\right) \cos(\theta_o) \\ &+\cos(g_e[t]) \cos(\phi_e) \sin(g_o[t])\Big) +\sin(\phi_e-\phi_o) \Big(\sin^2(\theta_e) \sin(\phi_e) \sin(g_o[t]) -2 \cos(g_e[t]) \cos(\phi_e) \sin^2\left(\frac{g_o[t]}{2}\right) \cos(\theta_o)\Big)\Bigg\},
\end{align*}
\begin{align*}
&\xi_{zy}(t) = \sin(\theta_e) \Bigg\{\cos(g_o[t]) \bigg[\sin(g_e[t]) \cos(\phi_e) \Big(\cos^2(\theta_o) \sin^2(\phi_o) +\cos^2(\phi_o) \Big) - \big(\cos(g_e[t])-1\big) \cos(\theta_e) \Big[\cos^2(\theta_o) \sin(\phi_o) \times  \\ & \times \cos(\phi_e-\phi_o) +\cos(\phi_o) \sin(\phi_e-\phi_o)\Big] +\sin(g_e[t]) \sin(\phi_e) \sin^2(\theta_o) \sin(\phi_o) \cos(\phi_o) \bigg] - \sin^2(\theta_o) \sin(\phi_o) \times \\ & \times \bigg[ \big(\cos(g_e[t])-1\big) \cos(\theta_e) \cos(\phi_e-\phi_o) +\sin(g_e[t]) \sin(\phi_e-\phi_o) \bigg] -\sin(g_o[t]) \cos(\theta_o) \bigg[ \Big(\cos(g_e[t])-1\Big) \cos(\theta_e) \cos(\phi_e) \\ & +\sin(g_e[t]) \sin(\phi_e) \bigg] \Bigg\},
\end{align*}
\begin{align*}
&\xi_{xz}(t) = \sin (\theta_o) \Bigg\{ \cos(g_e[t]) \Bigg[\cos(\phi_e-\phi_o) \bigg(2 \cos^2(\theta_e)
   \cos(\phi_e) \sin^2\left(\frac{g_o[t]}{2}\right) \cos(\theta_o) -\sin(\phi_e) \sin(g_o[t])\bigg) +\sin(\phi_e-\phi_o) \times \\ & \times \bigg( 2\sin(\phi_e) \sin^2\left(\frac{g_o[t]}{2}\right) \cos(\theta_o) + \cos^2(\theta_e) \cos(\phi_e) \sin(g_o[t]) \bigg) \Bigg] + \sin^2(\theta_e) \cos(\phi_e) \bigg(2 \sin^2\left(\frac{g_o[t]}{2}\right) \cos(\theta_o) \cos(\phi_e-\phi_o) \\ &+\sin(g_o[t]) \sin(\phi_e-\phi_o)\bigg) -\sin(g_e[t]) \cos(\theta_e) \bigg(2 \sin^2\left(\frac{g_o[t]}{2}\right) \cos(\theta_o) \sin(\phi_o) +\sin(g_o[t]) \cos(\phi_o) \bigg) \Bigg\},
\end{align*}
\begin{align*}
&\xi_{zx}(t) = \sin(\theta_e) \Bigg\{\sin^2(\theta_o) \cos(\phi_o) \bigg[2 \sin^2\left(\frac{g_e[t]}{2}\right) \cos (\theta_e) \cos(\phi_e-\phi_o) -\sin(g_e[t]) \sin(\phi_e-\phi_o) \bigg] -\cos(g_o[t]) \bigg[\sin(g_e[t]) \times \\ &\times \Big(\cos^2(\theta_o) \cos(\phi_o) \sin(\phi_e-\phi_o) +\sin(\phi_o) \cos(\phi_e-\phi_o)\Big) +\Big(\cos(g_e[t])-1\Big) \cos(\theta_e)\Big(\cos^2(\theta_o) \cos(\phi_o) \cos(\phi_e-\phi_o) \\ &-\sin(\phi_o) \sin(\phi_e-\phi_o) \Big) \bigg] -\sin(g_o[t]) \cos(\theta_o) \bigg[\sin(g_e[t]) \cos(\phi_e) -\Big(\cos(g_e[t])-1\Big) \cos(\theta_e) \sin(\phi_e) \bigg] \Bigg\},
\end{align*}
where we state explicitly the time-dependence of even and odd integral terms $g_e[t]$ and $g_o[t]$.

Once we have rotated the Hamiltonian into a suitable frame, the corresponding unitary operator for the evolution during a period $T$ can be rewritten using the Magnus expansion \cite{Bukov2015SM}:
\begin{equation}
\hat{\mathcal{U}}'(T) = \hat{\mathcal{T}} \exp \left( -i \int\limits_{0}^{T} dt' \hat{\mathcal{H}}'(t') \right) \approx \exp \left( -i \hat{\mathcal{H}}_{\mathrm{F}} T \right),
\end{equation}
with the Floquet Hamiltonian $\hat{\mathcal{H}}_{\mathrm{F}} =  \hat{\mathcal{H}}^{(0)}_{\mathrm{F}} + \hat{\mathcal{H}}^{(1)}_{\mathrm{F}} + \hat{\mathcal{H}}^{(2)}_{\mathrm{F}} + ... $, which consists of corrections in $\mathcal{O}[1/\omega]$. They can be written as 
\begin{align}
\label{eqS:H_F^0}
&\hat{\mathcal{H}}_{\mathrm{F}}^{(0)} = \frac{1}{T} \int\limits_{0}^{T} dt' \hat{\mathcal{H}}'(t'),\\
&\hat{\mathcal{H}}_{\mathrm{F}}^{(1)} = \frac{-i}{2! T} \int\limits_{0}^{T} dt' \int\limits_{0}^{t'} dt'' \big[ \hat{\mathcal{H}}'(t'), \hat{\mathcal{H}}'(t'')\big],\\
&\hat{\mathcal{H}}_{\mathrm{F}}^{(2)} = \frac{1}{3! T} \int\limits_{0}^{T} dt' \int\limits_{0}^{t'} dt'' \int\limits_{0}^{t''} dt''' \bigg\{ \Big[ \big[ \hat{\mathcal{H}}'(t'), \hat{\mathcal{H}}'(t'')\big], \hat{\mathcal{H}}'(t''') \Big] + \Big[ \big[ \hat{\mathcal{H}}'(t'''), \hat{\mathcal{H}}'(t'')\big], \hat{\mathcal{H}}'(t') \Big] \bigg\},
\end{align}
and higher-order terms can be written in a similar way using nested commutators. We notice that $||\hat{\mathcal{H}}_{\mathrm{F}}^{(k)} || \sim (1/\omega)^k$, and thus for very small time intervals, given by the period of the oscillating term $T = 2\pi/\omega \rightarrow 0$ (infinite frequency limit), the effective Floquet Hamiltonian is represented by the period-average of the time-dependent Hamiltonian written in Eq. (\ref{eqS:H_F^0}). In this work we consider the Floquet Hailtonian only to lowest order, $\hat{\mathcal{H}}_{\mathrm{F}} = \hat{\mathcal{H}}_{\mathrm{F}}^{(0)}$, while higher order corrections $\sim (1/\omega)^k$ ($k>0$) are accounted for in the numerical integration of the full time-dependent Hamiltonian.

Finally, let us choose the form of the oscillatoric magnetic field and find the period-averaged Floquet Hamiltonian for the transmon chain. This can be chosen in the form
\begin{equation}
\label{eqS:f_eo}
f_{\mathrm{e/o}}(t) = \frac{\lambda_{\mathrm{e/o}} \omega}{2 h_{\mathrm{e/o}}} \cos (\omega t + \varphi_{\mathrm{e/o}}), 
\end{equation}
where $\lambda_{\mathrm{e/o}}$ are constants of order unity, and $\varphi_{\mathrm{e/o}}$ are initial phases for the modulation. The latter is of high importance in the Floquet formalism, as it leads to kick-operator terms which change the basis of the system, but do not enter the effective time-independent Hamiltonian \cite{Goldman2014_SM}. However, in the current study we are interested in actual protocols with Floquet simulation, where the drive term is abruptly turned on at time point $t_0 = 0$, and the initial phase of the drive may be important. Here we consider zero initial phases $\varphi_{\mathrm{e/o}} = 0$, such that the kick operator is unity.

The integral functions $g_{\mathrm{e/o}}[t]$ are given by
\begin{equation}
\label{eqS:g_eo}
g_{\mathrm{e/o}}[t] = \lambda_{\mathrm{e/o}} \sin (\omega t). 
\end{equation}
Then, the period averaged coefficients $\overline{\xi_{\alpha \alpha'}} = (1/2\pi)\int_{0}^{2\pi}d\tau \xi_{\alpha \alpha'}(\tau)$ can be written in the form:
\begin{align*}
&\overline{\xi_{xx}} = \Bigg[\sin^2(\theta_e) \cos(\phi_e) \mathcal{J}_{0}(\lambda_o) \Big(\cos^2(\theta_o) \cos(\phi_o) \cos(\phi_e-\phi_o) -\sin(\phi_o) \sin (\phi_e-\phi_o)\Big) \\ &+\sin^2(\theta_o) \cos(\phi_o) \bigg[\mathcal{J}_{0}(\lambda_e) \Big(\cos^2(\theta_e) \cos(\phi_e) \cos(\phi_e-\phi_o) +\sin(\phi_e) \sin(\phi_e-\phi_o)\Big) +\sin^2(\theta_e) \cos(\phi_e) \cos(\phi_e-\phi_o)\bigg] \Bigg] \\ &+ \frac{\mathcal{J}_{0}(\lambda_e+\lambda_o)}{2} \bigg[\cos^2(\theta_e) \cos(\phi_e)
   \Big(\cos^2(\theta_o) \cos(\phi_o) \cos(\phi_e-\phi_o) -\sin(\phi_o) \sin(\phi_e-\phi_o)\Big) +\sin(\phi_e) \Big(\cos^2(\theta_o) \cos(\phi_o) \times \\ &\times \sin(\phi_e-\phi_o) +\sin(\phi_o) \cos(\phi_e-\phi_o)\Big) -\cos(\theta_e) \cos(\theta_o)\bigg] + \frac{\mathcal{J}_{0}(\lambda_e-\lambda_o)}{2} \bigg[\cos^2(\theta_e) \cos(\phi_e) \Big(\cos^2(\theta_o) \cos(\phi_o) \cos(\phi_e-\phi_o) \\ &-\sin(\phi_o) \sin(\phi_e-\phi_o)\Big)+\sin(\phi_e) \Big(\cos^2(\theta_o) \cos(\phi_o) \sin(\phi_e-\phi_o) +\sin(\phi_o) \cos(\phi_e-\phi_o)\Big) +\cos(\theta_e) \cos(\theta_o)\bigg],
\end{align*}
\begin{align*}
&\overline{\xi_{yy}} = \bigg\{\sin^2(\theta_e) \sin(\phi_e) \mathcal{J}_{0}(\lambda_o) \Big(\cos^2(\theta_o) \sin(\phi_o) \cos(\phi_e-\phi_o) +\cos(\phi_o) \sin(\phi_e-\phi_o)\Big) \\ &+\sin^2(\theta_o) \sin(\phi_o) \bigg[\sin(\phi_e) \cos(\phi_e-\phi_o) \Big(\cos^2(\theta_e) \mathcal{J}_{0}(\lambda_e) +\sin^2(\theta_e)\Big) -\mathcal{J}_{0}(\lambda_e) \cos(\phi_e) \sin(\phi_e-\phi_o)\bigg]\bigg\} \\ & +\frac{\mathcal{J}_{0}(\lambda_e+\lambda_o)}{2} \bigg\{\cos(\phi_e) \bigg[\cos(\phi_e) \Big(\cos^2(\theta_o) \sin^2(\phi_o) +\cos^2(\phi_o)\Big) +\sin(\phi_e) \sin^2(\theta_o) \sin(\phi_o) \cos(\phi_o)\bigg] \\ &+\cos^2(\theta_e) \sin(\phi_e) \Big(\cos^2(\theta_o) \sin(\phi_o) \cos(\phi_e-\phi_o)+ \cos(\phi_o) \sin(\phi_e-\phi_o)\Big) -\cos(\theta_e) \cos(\theta_o)\bigg\} \\ &+ \frac{\mathcal{J}_{0}(\lambda_e-\lambda_o)}{2} \bigg\{ \cos(\phi_e) \bigg[\cos(\phi_e) \Big(\cos^2(\theta_o) \sin^2(\phi_o) +\cos^2(\phi_o)\Big) +\sin(\phi_e) \sin^2(\theta_o) \sin(\phi_o) \cos(\phi_o)\bigg] \\ &+\cos^2(\theta_e) \sin(\phi_e) \Big(\cos^2(\theta_o) \sin(\phi_o) \cos(\phi_e-\phi_o) +\cos(\phi_o) \sin(\phi_e-\phi_o)\Big) +\cos(\theta_e) \cos(\theta_o) \bigg\},
\end{align*}
\begin{align*}
&\overline{\xi_{zz}} = \frac{1}{2} \sin(\theta_e) \sin(\theta_o) \cos(\phi_e-\phi_o) \Bigg\{2 \cos(\theta_e) \cos(\theta_o) \Big(1 - \mathcal{J}_{0}(\lambda_e) -\mathcal{J}_{0}(\lambda_o)\Big) +\Big(\cos(\theta_e) \cos(\theta_o)-1\Big) \mathcal{J}_{0}(\lambda_e+\lambda_o) \\ &+\Big(\cos(\theta_e) \cos(\theta_o)+1\Big) \mathcal{J}_{0}(\lambda_e-\lambda_o)\Bigg\},
\end{align*}
\begin{align*}
&\overline{\xi_{xy}} = \frac{1}{16} \Bigg\{\sin^2(\theta_o) \sin(2\phi_o) \bigg[2 \sin^2(\theta_e) \Big[2 -4 \cos^2(\phi_e) \mathcal{J}_{0}(\lambda_o) +\cos(2\phi_e) \Big(\mathcal{J}_{0}(\lambda_e-\lambda_o)+\mathcal{J}_{0}(\lambda_e+\lambda_o) -2\mathcal{J}_{0}(\lambda_e)+2\Big) \Big] \\ & +\Big(\cos(2\theta_e)+3\Big) \Big(2\mathcal{J}_{0}(\lambda_e) -\mathcal{J}_{0}(\lambda_e-\lambda_o)-\mathcal{J}_{0}(\lambda_e+\lambda_o) \Big) \bigg] +\sin^2(\theta_e) \sin(2 \phi_e) \bigg[\Big(2\mathcal{J}_{0}(\lambda_o) -\mathcal{J}_{0}(\lambda_e-\lambda_o) \\ & -\mathcal{J}_{0}(\lambda_e+\lambda_o) \Big) \Big(2 \sin^2(\theta_o) \cos(2\phi_o) +\cos(2\theta_o)+3\Big) -8 \Big(\mathcal{J}_{0}(\lambda _e)-1\Big) \sin^2(\theta_o) \sin^2(\phi_o) \bigg]\Bigg\},
\end{align*}
\begin{align*}
&\overline{\xi_{yx}} = \frac{1}{2} \Bigg\{\sin^2(\theta_e) \bigg[ \Big(1-\mathcal{J}_{0}(\lambda_e)\Big) \sin(2\phi_e) \sin^2(\theta_o) \cos^2(\phi_o) +2 \sin(\phi_e) \mathcal{J}_{0}(\lambda_o) \Big(\cos^2(\theta_o) \cos(\phi_o) \cos(\phi_e-\phi_o) \\ & -\sin(\phi_o) \sin(\phi_e-\phi_o)\Big) \bigg] +\sin^2(\theta_o) \sin(2\phi_o) \bigg[ \sin^2(\phi_e) \Big(\cos^2(\theta_e) \mathcal{J}_{0}(\lambda_e) +\sin^2(\theta_e) \Big) +\mathcal{J}_{0}(\lambda_e) \cos^2(\phi_e) \bigg] \\ & +\mathcal{J}_{0}(\lambda_e-\lambda_o) \bigg[ \cos^2(\theta_e) \sin(\phi_e) \Big(\cos^2(\theta_o) \cos(\phi_o) \cos(\phi_e-\phi_o)-\sin(\phi_o) \sin(\phi_e-\phi_o)\Big) -\cos(\phi_e) \times \\ & \times \Big(\cos^2(\theta_o) \cos(\phi_o) \sin(\phi_e-\phi_o)+\sin(\phi_o) \cos(\phi_e-\phi_o)\Big) \bigg] +\mathcal{J}_{0}(\lambda_e+\lambda_o) \bigg[\cos^2(\theta_e) \sin(\phi_e) \Big(\cos^2(\theta_o) \cos(\phi_o) \cos(\phi_e-\phi_o) \\ & -\sin(\phi_o) \sin(\phi_e-\phi_o)\Big)-\cos(\phi_e) \Big(\cos^2(\theta_o) \cos(\phi_o) \sin(\phi_e-\phi_o) +\sin(\phi_o) \cos(\phi_e-\phi_o)\Big) \bigg] \Bigg\},
\end{align*}
\begin{align*}
&\overline{\xi_{yz}} = \frac{1}{2} \Bigg\{\sin(2\theta_o) \bigg[ \sin(\phi_e) \cos(\phi_e-\phi_o) \Big[\cos^2(\theta_e) \mathcal{J}_{0}(\lambda_e) -\sin^2(\theta_e) \Big(\mathcal{J}_{0}(\lambda_o)-1\Big) \Big] -\mathcal{J}_{0}(\lambda_e) \cos(\phi_e) \sin(\phi_e-\phi_o) \bigg] \\ & -\sin(\theta_o) \mathcal{J}_{0}(\lambda_e-\lambda_o) \Big(\cos^2(\theta_e) \sin(\phi_e) \cos(\theta_o) \cos(\phi_e-\phi_o) +\cos(\theta_e) \sin(\phi_o) -\cos(\phi_e) \cos(\theta_o) \sin(\phi_e-\phi_o) \Big) \\ & +\sin(\theta_o) \mathcal{J}_{0}(\lambda_e+\lambda_o) \bigg[\cos(\theta_e) \sin(\phi_o) +\cos(\phi_e) \cos(\theta_o) \sin(\phi_e-\phi_o) - \cos^2(\theta_e) \sin(\phi_e) \cos(\theta_o) \cos(\phi_e-\phi_o) \bigg] \Bigg\},
\end{align*}
\begin{align*}
&\overline{\xi_{zy}} = \frac{1}{2} \sin(\theta_e) \Bigg\{ -\cos(\theta_e) \Big(\mathcal{J}_{0}(\lambda_e-\lambda_o) +\mathcal{J}_{0}(\lambda_e+\lambda_o)  -2 \mathcal{J}_{0}(\lambda_o)\Big) \Big(\cos^2(\theta_o) \sin(\phi_o) \cos(\phi_e-\phi_o) +\cos(\phi_o) \sin(\phi_e-\phi_o)\Big) \\ & -2 \cos(\theta_e) \Big(\mathcal{J}_{0}(\lambda_e)-1\Big) \sin^2(\theta_o) \sin(\phi_o) \cos(\phi_e-\phi_o) +\sin(\phi_e) \cos(\theta_o) \Big(\mathcal{J}_{0}(\lambda_e+\lambda_o)-\mathcal{J}_{0}(\lambda_e-\lambda_o)\Big) \Bigg\},
\end{align*}
\begin{align*}
&\overline{\xi_{xz}} = \frac{1}{2} \Bigg\{ \sin(2\theta_o) \Bigg[ \cos(\phi_e) \cos(\phi_e-\phi_o) \bigg[\cos^2(\theta_e) \mathcal{J}_{0}(\lambda_e) -\sin^2(\theta_e) \Big(\mathcal{J}_{0}(\lambda_o) -1\Big)\bigg] + \mathcal{J}_{0}(\lambda_e) \sin(\phi_e) \sin(\phi_e-\phi_o) \Bigg] \\ & -\sin(\theta_o) \mathcal{J}_{0}(\lambda_e+\lambda_o) \Big(\cos^2(\theta_e) \cos(\phi_e) \cos(\theta_o) \cos(\phi_e-\phi_o) -\cos(\theta_e) \cos(\phi_o) +\sin(\phi_e) \cos(\theta_o) \sin(\phi_e-\phi_o)\Big) \\ & -\sin(\theta_o) \mathcal{J}_{0}(\lambda_e-\lambda_o) \bigg[\cos(\theta_e) \Big(\cos(\theta_e) \cos(\phi_e) \cos(\theta_o) \cos(\phi_e-\phi_o) +\cos(\phi_o)\Big) +\sin(\phi_e) \cos(\theta_o) \sin(\phi_e-\phi_o) \bigg] \Bigg\},
\end{align*}
\begin{align*}
&\overline{\xi_{zx}} = \frac{1}{2} \Bigg\{ \sin(2\theta_e) \Bigg[\cos(\phi_o) \cos(\phi_e-\phi_o) \bigg[\cos^2(\theta_o) \mathcal{J}_{0}(\lambda_o) -\Big(\mathcal{J}_{0}(\lambda_e)-1\Big) \sin^2(\theta_o)\bigg] -\mathcal{J}_{0}(\lambda_o) \sin(\phi_o) \sin(\phi_e-\phi_o)\Bigg] \\ & +\sin(\theta_e) \mathcal{J}_{0}(\lambda_e+\lambda_o) \Big( \cos(\phi_e) \cos(\theta_o) +\cos(\theta_e) \sin(\phi_o) \sin(\phi_e-\phi_o) -\cos(\theta_e) \cos^2(\theta_o) \cos(\phi_o) \cos(\phi_e-\phi_o) \Big) \\ & -\sin(\theta_e) \mathcal{J}_{0}(\lambda_e-\lambda_o)  \bigg[ \cos(\theta_o) \Big(\cos(\theta_e) \cos(\theta_o) \cos(\phi_o) \cos(\phi_e-\phi_o) +\cos(\phi_e)\Big) -\cos(\theta_e) \sin(\phi_o) \sin(\phi_e-\phi_o) \bigg] \Bigg\}.
\end{align*}
The above equations define the exact form of the generic Hamiltonian (4) from the main text, and thus describe the possible Hamiltonians accessible for the Floquet quantum simulation with this method, assuming different even/odd periodic cosine modulation. 

\subsection{\label{sect:B}Transverse Ising model derivation and intuitive explanation}

\begin{figure}[h!]
\includegraphics[width=0.6\linewidth]{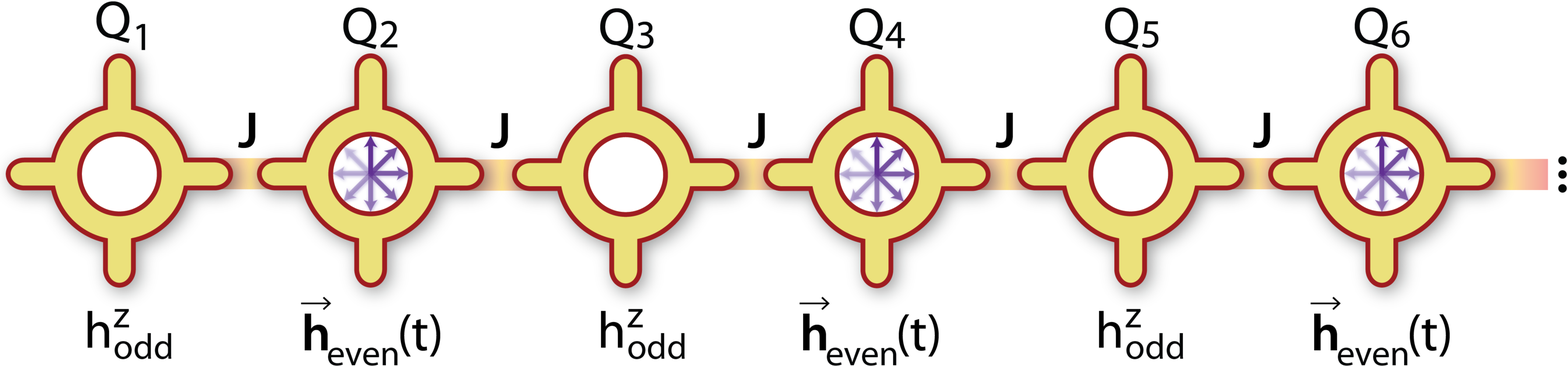}
\caption{Superconducting qubit chain with isotropic XY interaction $J$, static effective magnetic field in $z$ direction on the odd sublattice, $h^z_{\mathrm{odd}}$, and fast time-dependent magnetic field acting on even sublattice sites, $\mathbf{h}_{\mathrm{even}}(t)$.}
\label{figS:sketch_Ising}
\end{figure}

In this section we provide a detailed procedure to engineer the transverse Ising Hamiltonian as an effective Floquet Hamiltonian of the isotropic XY model with transverse and longitudinal fields. For this we consider a system with two (odd and even) sublattices, where only one of the sublattices experience fast oscillations of the magnetic field (see sketch in Fig. \ref{figS:sketch_Ising}).

The intuitive way to describe the realization of the Ising model ($\propto \sigma_1^x \sigma_2^x$) from the isotropic XY case (with interaction type $\propto \sigma_1^x \sigma_2^x + \sigma_1^y \sigma_2^y$) is to take two spins and imagine one to be rotating in a magnetic field. Starting from the flip-flop interaction, if we choose the axis of the magnetic field to be in the $\mathbf{e}^x$ direction, nothing will happen to the first interaction term. At the same time, the rotation of the second spin leads to oscillations of the second term between $\pm \sigma_1^y \sigma_2^y$, and also induce a $\propto \sigma_1^y \sigma_2^z$ interaction component. For large frequencies of oscillation and carefully chosen drive amplitude, the plus and minus components will cancel each other, as well as the cross-interaction components, ultimately leaving the Ising term $\propto \sigma_1^x \sigma_2^x$ only.

Having in mind the aforementioned intuitive explanation, and taking the full solution from section A, the starting Hamiltonian for the simulation of the Ising model in the transmon chain reads:
\begin{align}
\label{eqS:H_XY_X+Z}
&\hat{\mathcal{H}}(t) = J \sum\limits_{j=1}^{N-1} (\sigma_j^x \sigma_{j+1}^x + \sigma_j^y \sigma_{j+1}^y) + \sum\limits_{j=1}^{\lfloor N/2 \rfloor} \lambda \omega \cos(\omega t) \sigma^x_{2j} + \hat{\mathcal{H}}_{\mathrm{magn}}(t),
\end{align}
where $\lambda (\equiv \lambda_{\mathrm{even}})$ is a drive parameter, and $\mathcal{J}_0[x]$ denotes the zeroth order Bessel function of the first kind. In the infinite frequency limit $|J|/\omega \rightarrow 0$ and for $\lambda = 1.20241$ (such that $\mathcal{J}_0[2 \lambda] = 0$) the interaction term can be reduced to the Ising type. Additionally, the last term in Eq. (\ref{eqS:H_XY_X+Z}) is designed to introduce a transverse effective magnetic field and can be written as:
\begin{align}
\label{eqS:H_magn}
&\hat{\mathcal{H}}_{\mathrm{magn}}(t) = \sum\limits_{j=1}^{\lceil N/2 \rceil} h^z \sigma^z_{2j-1} + \frac{2 h^z}{1+\mathcal{J}_0[4 \lambda]} \sum\limits_{j=1}^{\lfloor N/2 \rfloor} \cos(2 \lambda \sin[\omega t]) \sigma^z_{2j}.
\end{align}
The first term in Eq. (\ref{eqS:H_magn}) is a static magnetic field on the odd sublattice, and commutes trivially with the fast oscillation part. However, the magnetic field on the even sublattice can be modified by the drive. The second term representing the magnetic field, deviates from the general form considered in Sec. \ref{sect:A}, since it does not have the same time dependence. As opposed to the other field, however, the magnitude of this field does not increase with increasing $\omega$ and can thus be treated as a perturbation. Since it doesn't commute with the main driving field, it will be strongly modified by it.  Going to the rotating frame with the unitary operator $\hat{\mathcal{U}}_R(t) = \exp\left\{ i \lambda \sin(\omega t) \right\}$, the magnetic term becomes
\begin{align}
\label{eqS:H_magn_rotated}
&\hat{\mathcal{H}}'_{\mathrm{magn}}(t) = h^z \sum\limits_{j=1}^{\lceil N/2 \rceil} \sigma^z_{2j-1} + \frac{2 h^z}{1+\mathcal{J}_0[4 \lambda]} \cos(2 \lambda \sin[\omega t]) \sum\limits_{j=1}^{\lfloor N/2 \rfloor} \left( \cos(2 \lambda \sin(\omega t)) \sigma^z_{2j} + \sin(2 \lambda \sin(\omega t)) \sigma^y_{2j}\right).
\end{align}
Finally, performing the period averaging, the $\sigma^y_{2j}$ term vanishes trivially since the sine function oscillates between positive and negative values. At the same time, given that $\lambda$ is fixed by the condition $\mathcal{J}_0[2 \lambda] = 0$, the integral $\int_{0}^{2\pi} dx \cos^2(2 \lambda \sin[x]) = \pi (1 + \mathcal{J}_0[4 \lambda])$ gives a finite result
\begin{equation}
\label{eqS:H_magn_F}
\hat{\mathcal{H}}_{\mathrm{F}}^{\mathrm{magn}} = h^z \sum\limits_{j=1}^{N} \sigma^z_{j},
\end{equation}
thus allowing to introduce an effective transverse field $h^z$. 

\subsection{\label{sect:C}Digital simulation of transverse Ising model}

\begin{figure}[h!]
\includegraphics[width=0.9\linewidth]{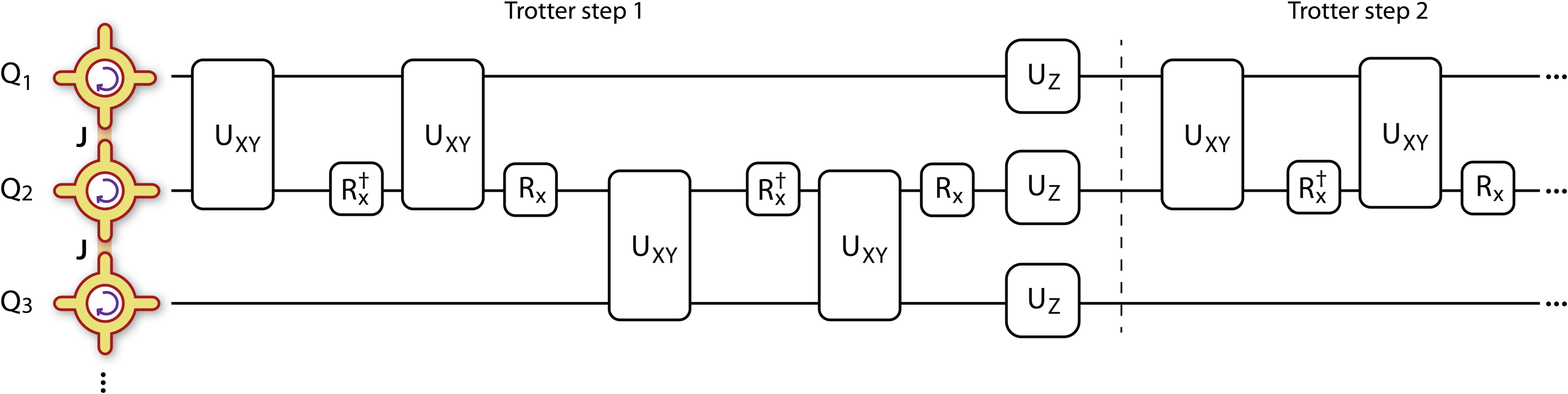}
\caption{Digital simulation scheme from Ref. \cite{LasHeras2014_SM}. It relies on Trotterizaton of the isotropic XY model dynamics, where additional single qubit rotations at every second site effectively eliminates the YY coupling.}
\label{figS:digital}
\end{figure}

In this section we describe the digital simulation protocol, which we use to benchmark the performance of the Floquet quantum simulator. The transverse Ising model simulation, which we will consider, was theoretically described in Ref. \cite{LasHeras2014_SM} and experimentally realized in Ref. \cite{Salathe2015_SM}. The circuit scheme is shown in Fig. \ref{figS:digital}.

The algorithm relies on the preparation of a unitary transformation with the effective Hamiltonian of interest using repetitions of a small step, corresponding to the Trotterization procedure. The protocol for the simulation of an arbitrary k-local Hamiltonian $\hat{\mathcal{H}}$ (generally not available in the physical setup) relies on the sequential implementation of the available parts of a Hamiltonian $\hat{\mathcal{H}}_k$ (constructed from gates) such that $\sum_k \hat{\mathcal{H}}_k = \hat{\mathcal{H}}$. The corresponding unitary of a single digital step $j$ of duration $\delta t$ reads
\begin{equation}
\label{eqS:U_Trot_1}
\hat{U}_{j} (\delta t)= e^{-i\hat{\mathcal{H}}_1 \delta t} e^{-i\hat{\mathcal{H}}_2 \delta t} ... e^{-i\hat{\mathcal{H}}_k \delta t},
\end{equation}
and the implementation of $N_{\mathrm{Tr}} \rightarrow \infty$ Trotter steps combine into the unitary $\hat{U}(t) = \lim_{N_{\mathrm{Tr}}\rightarrow \infty}\hat{U}_j(\delta t)^{N_{\mathrm{Tr}}} \approx e^{-i\hat{\mathcal{H}} t}$.

In this spirit, the implementation of the transverse Ising model model was proposed to rely on multiple applications of the Trotter step graphically shown in Fig. \ref{figS:digital}. It starts with the implementation of $U_{\mathrm{XY}} = \exp\{-i \delta t \sum\limits_{j=1}^{N/2} \hat{\mathcal{H}}^{(+)}_{2j-1,2j} \}$, where by $\hat{\mathcal{H}}^{(+)}_{j,j'} = \frac{J}{2} (\sigma^x_{j} \sigma^x_{j'} + \sigma^y_{j} \sigma^y_{j'})$ we define the simple application of XY interaction for each pair of qubits. Next, this unitary can be rotated by applying $\pi$ X rotations at every second qubit, $\hat{R}_x = \exp \{-i \frac{\pi}{2} \sum_{j=1}^{N/2} \sigma^x_{2j} \}$, which leads to $\hat{R}_x^\dagger U_{\mathrm{XY}} \hat{R}_x = \exp\{-i \delta t \sum\limits_{j=1}^{N/2} \hat{\mathcal{H}}^{(-)}_{2j-1,2j} \}$, where we define $\hat{\mathcal{H}}^{(-)}_{j,j'} = \frac{J}{2} (\sigma^x_{j} \sigma^x_{j'} - \sigma^y_{j} \sigma^y_{j'})$ as the XY Hamiltonian with the YY term flipped by rotation. Finally, the last layer in the Trotter step implements the transverse fields with $U_{\mathrm{Z}} = \exp\{-i \delta t h^z \sum\limits_{j=1}^{N} \sigma_j^z \}$. Once the Trotter step is repeated many times, the non-commuting Hamiltonian parts can be added, thus implementing the transverse Ising model digitally.

The same considerations can be repeated for the digitized annealing procedure \cite{Barends2016_SM}. Here the important part is to keep the phase applied by $U_{\mathrm{Z}}$ gates consistent with the adiabatic evolution.

\subsection{\label{sect:D}Accounting for the finite anharmonicity}

\begin{figure}
\includegraphics[width=0.95\linewidth]{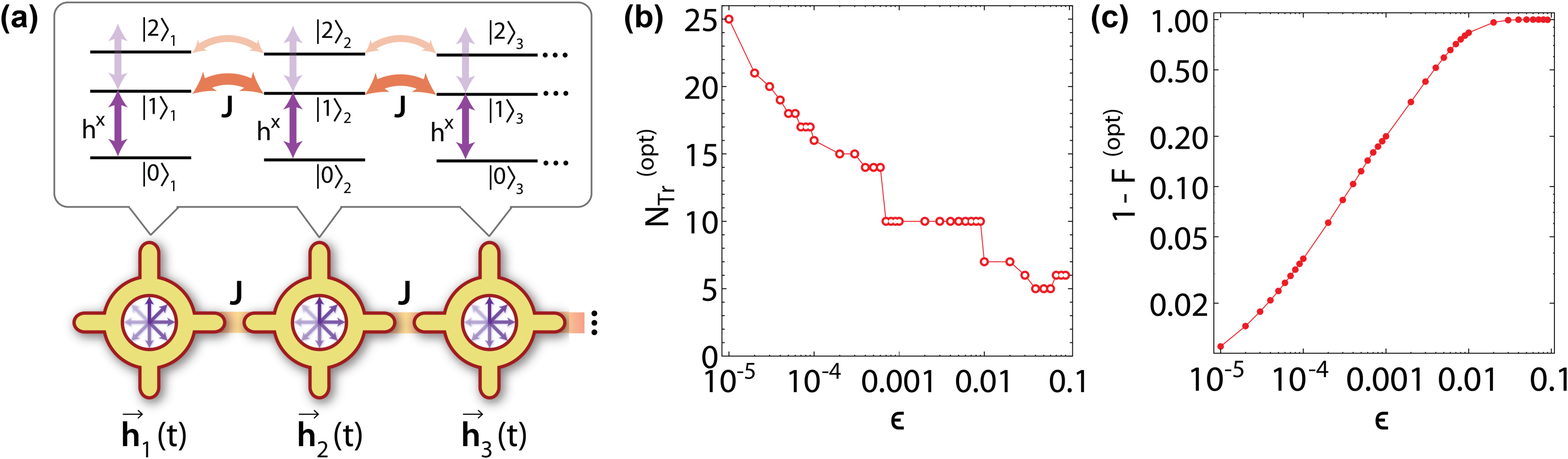}
\caption{\textbf{Accounting for finite anharmonicity.} (a) Sketch of a realistic transmon chain with weakly anharmonic multilevel structure. In each circuit we account for $\{ |0\rangle_j, |1\rangle_j, |2\rangle_j \}$ states. The infidelity of the simulation arises from microwave driving of the $|1\rangle_j \leftrightarrow |2\rangle_j$ transition, and additional flip-flop coupling. (b) Optimal number of the Trotter steps $N_{\mathrm{Tr}}^{\mathrm{(opt)}}$ plotted for different values of the gate error. (c) Corresponding optimal infidelity as a function of the gate error. The results are shown for transverse Ising annealing with $t_f = 15 |J|^{-1}$ and $N = 4$.}
\label{figS:leakage}
\end{figure}

In this section we consider the transmon chain Hamiltonian and account for finite anharmonicity. This can be described by the Hamiltonian
\begin{align}
\label{eqS:H_anharm}
\hat{\mathcal{H}} = 2 J \sum\limits_{j=1}^{N-1}(\hat{a}_{j}^\dagger \hat{a}_{j+1} + \hat{a}_{j} \hat{a}_{j+1}^\dagger) + \sum\limits_{j=1}^{N} \left\{ \Delta_j \hat{a}_j^\dagger \hat{a}_j + (\Omega_j \hat{a}_j + \Omega_j^* \hat{a}_j^\dagger) \right\} + \sum\limits_{j=1}^{N} \frac{A}{2} \hat{a}_j^\dagger \hat{a}_j^\dagger \hat{a}_j \hat{a}_j,
\end{align}
where $\hat{a}_j^\dagger$ ($\hat{a}_j$) corresponds to the creation (annihilation) operator for excitations of the $j$-th transmon circuit. The first term in Eq. (\ref{eqS:H_anharm}) corresponds to nearest-neighbour capacitive coupling for transmons. The second term in curly brackets denotes an effective magnetic field in the $z$ direction given by the flux-bias dependent detuning $\Delta_j$ and the microwave drive terms $\Omega_j$ corresponding to an effective magnetic field in $xy$ plane. The last term of Eq.~(\ref{eqS:H_anharm}) corresponds to the anharmonicity of the circuit $A$ provided by Josephson junctions. In the case of infinitely large anharmonicity the Hamiltonian (\ref{eqS:H_anharm}) can be projected onto the lowest occupation subspace for each qubit $\{ |0\rangle, |1\rangle \}_j$, accounting only for singly excited circuits. This allows for a spin-1/2 description of the chain, and subsequent simulation of quantum magnetism. However, in realistic transmon samples the anharmonicity is typically small, and higher states of the circuit must be accounted for [see sketch in Fig. \ref{figS:leakage}(a)]. In particular, this is important for the case of a strong microwave drive $\Omega$, as it leads to non-zero occupation of higher lying states, corresponding to the leakage of information out of the logical subspace. This can largely decrease the fidelity, and typically is the bottleneck for fast digital computation.

In this study we consider the effects of finite $A$ by expanding the Hilbert space for each site to have doubly occupied states, $\{ |0\rangle, |1\rangle, |2\rangle \}_j$. As a test case we take the ground state preparation of the transverse Ising model, studied for the case of infinite anharmonicity in the preceding sections. The annealing schedule is chosen in the same form, making use of the $h^x \leftrightarrow \Omega (\Omega^*)$ and $h^z \leftrightarrow 2 \Delta$ correspondence.

The main results of the finite $A$ scaling for the Floquet quantum simulation is presented in Fig. 4 and the corresponding section of the main text. Here we provide more details of the Floquet to digital benchmarking procedure. The estimate of the digital protocol infidelity accounts for several contributions. The first comes from the Trotterization procedure $\epsilon_{\mathrm{dig}}(N_{\mathrm{Tr}})$ and depends strongly on the number of Trotter steps, favouring long sequences. The second contribution is a total infidelity from gate operations $\epsilon_{\mathrm{gates}}(N_{\mathrm{Tr}}) = 1 - \left[1 - (5 N - 4) \epsilon \right]^{N_{\mathrm{Tr}}}$, which increases with the number of Trotter steps. The optimization procedure is performed for different values of the gate errors $\epsilon$. The results are plotted in Fig. \ref{figS:leakage}(b), and the optimal Trotter step numbers $N_{\mathrm{Tr}}^{\mathrm{opt}}$ is shown to decrease in the case of large gate errors. The corresponding optimal infidelity [Fig. \ref{figS:leakage}(c)] shows a significant increase for $\epsilon > 10^{-3}$. 

To compare the digital and Floquet approaches, we should compare how each of the two approaches could be implemented on comparable physical systems. To this end we consider transmons with the same anharmonicity $A$, and assume that they also have comparable decoherence rates (but note that we assume that the physical coupling $J$ can be different in the two scenarios). Therefore, to have a similar influence of decoherence in the two approaches we assume that the simulations need to be completed in the same time. To simulate the same evolution this requires that the two approaches have the same $J_{\mathrm{sim}}$ and thus the same values of $A/J_{\mathrm{sim}}$. In the Floquet case this is defined by the $A/\omega$ and $\omega/J$ ratios. For the digital simulation, the relation is more subtle and relies on the scaling of the gate time with $A$ for a fixed error value. The full discussion of this complex subject lies beyond the scope of present study, and for simplicity we just assume that each gate can be implemented in a time $t_{\mathrm{gate}} = c/A$, where $c$ is a constant which controls the quality of the gate. Taking the existing studies \cite{Rol2016_SM,Motzoi2009_SM,Martinis2014_SM}, and considering a best case scenario, we set $c=35$. For realistic devices with $A=2\pi \times 300$ MHz this will correspond to 18 ns gates, and we will consider a low error of $\epsilon = 10^{-4}$. This will be extremely challenging to achieve with current technologies, but for the example considered in the main text this is what is required to achieve a performance comparable to what is achievable with the Floquet approach. With more realistic numbers the performance of the digital approach will be less ideal and thus not comparable to the result which can be achieved with the Floquet simulation.

\subsection{\label{sect:E}Floquet simulation of spin-1/2 XYZ model}

To simulate the XYZ model in the Floquet basis, we start with the time-dependent Hamiltonian in the form 
\begin{equation}
\label{eq:H_XX+h}
\hat{\mathcal{H}}(t) = \sum\limits_{j=1}^{N-1} J (\sigma^x_{j} \sigma^x_{j+1} + \sigma^y_{j} \sigma^y_{j+1}) + \sum\limits_{j=1}^{N} \mathbf{h}(t) \cdot \boldsymbol{\sigma}_j + \hat{\mathcal{H}}_{z},
\end{equation}
where we consider the oscillating effective magnetic field to be homogeneous for all sites. $\hat{\mathcal{H}}_{z}$ describes the part of the Hamiltonian responsible for implementing the static $z$-oriented magnetic field for annealing. Considering $\mathbf{h}(t) = \lambda \omega \cos(\omega t) \mathbf{e}^x$ with $\omega$ being the largest energy scale, and setting $\lambda = 1.81144$, we eliminate the cross-terms and are left with a Floquet Hamiltonian of the form
\begin{equation}
\label{eq:H_XYZ}
\hat{\mathcal{H}}_{F} = \sum\limits_{j=1}^{N-1} (J^x \sigma_j^x \sigma_{j+1}^x + J^y \sigma_j^y \sigma_{j+1}^y + J^z \sigma_j^z \sigma_{j+1}^z) + h^z \sum\limits_{j=1}^{N} \sigma_j^z \equiv \hat{\mathcal{H}}_{\mathrm{XYZ}},
\end{equation}
where the couplings are $J^y = 2 J^x/ 3$, $J^z = J^x/3$, and $J^x = J < 0$. Here the simulated coupling changes for the YY and ZZ interaction components, and we consider $J^x = J_{\mathrm{sim}}$ as a reference. We note that as compared to the transverse Ising case this Hamiltonian possesses a small energy gap, and in the absence of an additional transverse field it is difficult to anneal even with the ideal continuous Hamiltonian.
\begin{figure}[t]
\includegraphics[width=0.55\linewidth]{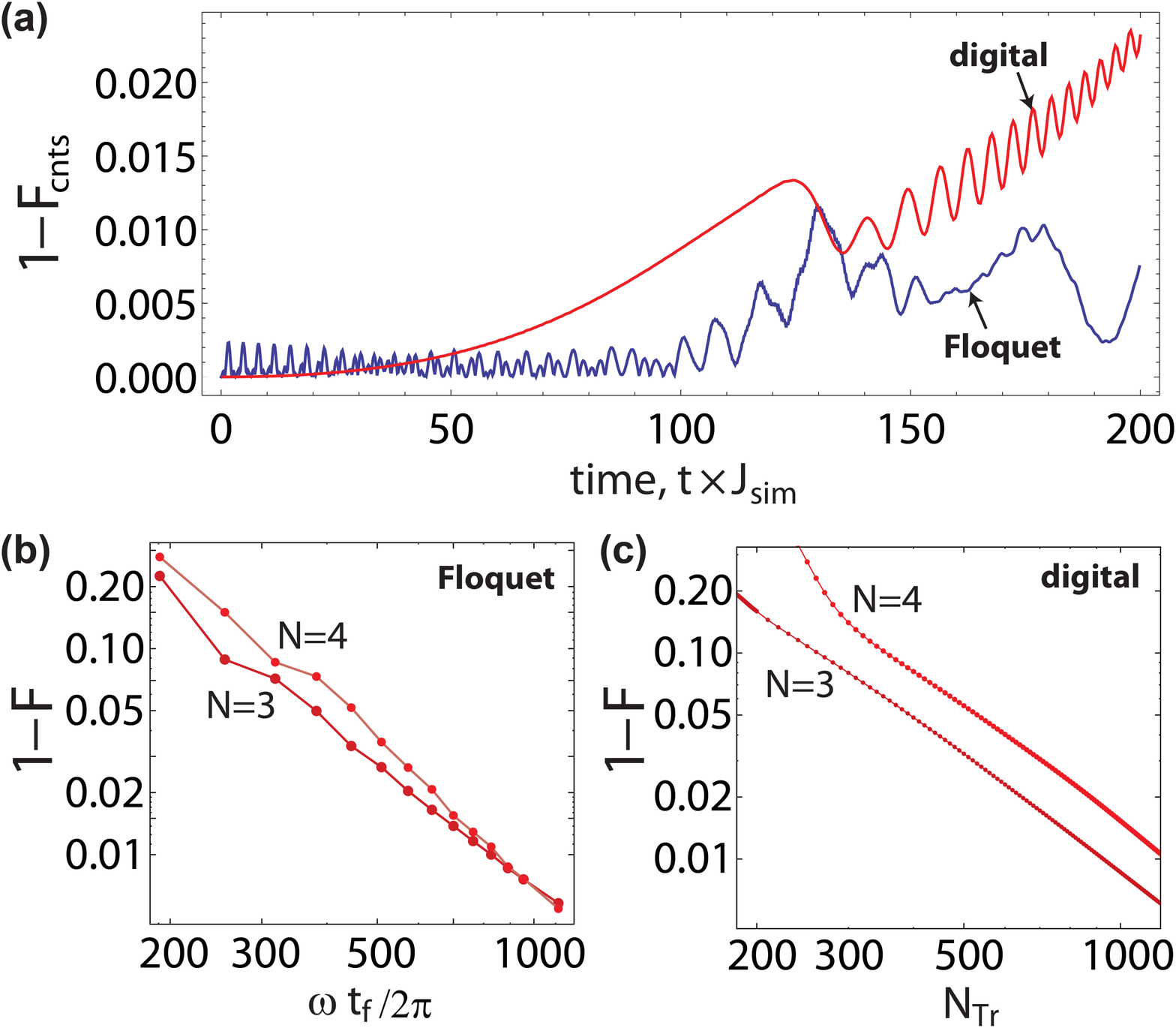}
\caption{\textbf{XYZ annealing.} (a) Time dependence of the infidelity for the Floquet evolution (blue), and optimal digital annealing (red). Here the infidelity is measured with respect to the instantaneous wavefunction of the continuous annealing. The digital evolution contains $N_{\mathrm{Tr}} = 477$ Trotter steps, equal to half the number of stroboscopic periods. (b) Modulation frequency dependence of the final state infidelity for $t_f = 200 |J|^{-1}$ annealing, measured with respect to an ideal target state. (c) Final infidelity of the digital quantum simulation of the XYZ model as a function of the number of Trotter steps. $F$ is measured with respect to an ideal target state.}
\label{figS:non-stq-annl}
\end{figure}

To characterize the Floquet and digital simulation procedures we plot the instantaneous infidelity with respect to the continuous annealing case, observing how closely one can follow the ground state [Fig. \ref{figS:non-stq-annl}(a)]. The blue curve for the Floquet simulation at stroboscopic times shows that the deviation begins to grow once we approach the critical point. To compare with the digital procedure, we plot the infidelity for the Trotterization approach. We assume the number of Trotter steps is equal to half of number of Floquet periods, $N_{\mathrm{Tr}} = 1/2(t_f/T) = 477$ (see below for the details of the circuit). This will be an upper bound for the number of Trotter steps for the digital simulator for the same resources. This can be deduced from the digital simulation protocol assuming that the time required to implement the two-qubit gate is inversely proportional to the coupling strength, $t_{\mathrm{gate}} \sim |J|^{-1}$, and in addition the gates on the two sublattices have to be applied separately. We note that considering different ordering of the gates results in largely different results for the final state infidelity. While we have not performed a full optimization of this ordering, the results presented in the figure is the result of the optimization over 24 different possibilities for a Trotter step composition. This suggests that the digital procedure is strongly model dependent, and extra resources are required for sequence optimization \cite{Wecker2015_SM,Reiher2016_SM}. 
In Fig. \ref{figS:non-stq-annl}(b) we plot the dependence of the Floquet XYZ annealing on the drive frequency, showing the resources necessary for high fidelity annealing as a function of the number of stroboscopic periods. A similar analysis is performed for the digital procedure, where the dependence of Trotter step number is considered [Fig. \ref{figS:non-stq-annl}(c)]. Akin to the transverse Ising annealing case described in the main text, the Floquet approach shows smaller infidelity for limited number of steps, in particular for $N=4$. Unlike the transverse Ising model, however even for $N_{\mathrm{Tr}}$ as large as thousand steps, the digital approach fails provide smaller infidelity as compared to Floquet approach, but this may change if even higher number of Trotter steps are considered.

\begin{figure}
\includegraphics[width=0.95\linewidth]{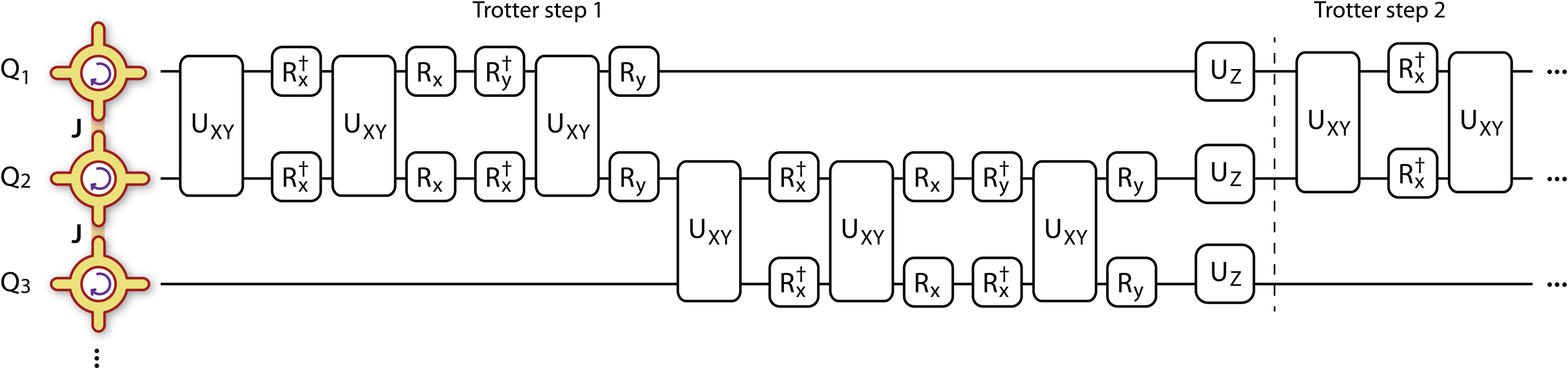}
\caption{Digital simulation scheme of the Heisenberg-type model from Ref. \cite{LasHeras2014_SM}. It relies on the sequential application of bare XY couplings ($\propto \alpha_{xy} J(\sigma^x_j \sigma^x_{j+1} + \sigma^y_j \sigma^y_{j+1})$), and its $\pi/2$ rotated version corresponding to XZ ($\propto \alpha_{xz} J(\sigma^x_j \sigma^x_{j+1} + \sigma^z_j \sigma^z_{j+1})$) and YZ ($\propto \alpha_{yz} J(\sigma^y_j \sigma^y_{j+1} + \sigma^z_j \sigma^z_{j+1})$) interactions. Here $\alpha_{xy,xz,yz}$ correspond to dimensionless coefficients which allow tuning the model into the anisotropic XYZ model. We note that the described sequence can be further optimized by combing single qubit rotations to unity matrices and $z$ rotations. Moreover, the ordering of the terms can be changed to optimize the procedure.}
\label{figS:digital_XYZ}
\end{figure}

\textit{Digital XYZ model.---}The considered digital simulation protocol, originally described in Ref. \cite{LasHeras2014_SM}, is sketched in Fig. \ref{figS:digital_XYZ}.
It relies on the sequential rotation of the basis for nearest-neighbour interaction, such that in the limit of a large number of Trotter steps it sums up to $\sum_{j=1}^{N-1} (J^x \sigma_j^x \sigma_{j+1}^x + J^y \sigma_j^y \sigma_{j+1}^y + J^z \sigma_j^z \sigma_{j+1}^z)$. First, the XY unitary is performed, implementing $U_{\mathrm{XY}} = \exp\{-i \delta t \sum\limits_{j=1}^{N-1} \alpha_{xy} J(\sigma^x_j \sigma^x_{j+1} + \sigma^y_j \sigma^y_{j+1}) \}$, where $\alpha_{xy}$ is some constant. Next, applying $\pi/2$ rotation around the $x$ axis for each qubit, $\hat{R}_x = \exp \{-i \frac{\pi}{4} \sum_{j=1}^{N} \sigma^x_{j} \}$, the two-qubit unitary can be transformed to $U_{\mathrm{XZ}} = \exp\{-i \delta t \sum\limits_{j=1}^{N-1} \alpha_{xz} J (\sigma^x_j \sigma^x_{j+1} + \sigma^z_j \sigma^z_{j+1}) \}$. Subsequent $\pi/2$ rotation around the $y$ axis will implement the $U_{\mathrm{YZ}} = \exp\{-i \delta t \sum\limits_{j=1}^{N/2} \alpha_{yz} J (\sigma^y_j \sigma^y_{j+1} + \sigma^z_j \sigma^z_{j+1}) \}$ interaction. For instance, the final chosen configuration of $J_y = 2 J^x/ 3$, $J^z = J^x/3$ can be achieved by choosing $J_x = J$ (can be different from $J_{\mathrm{sim}}$), $\alpha_{xy} = 2/3$, $\alpha_{xz}=1/3$, $\alpha_{yz}=0$, simplifying the gate sequence. Finally, the $U_{\mathrm{Z}}$ operation introduces an effective magnetic field in the $z$ direction which allows for annealing to the ground state of the XYZ model. The linear schedule can then be designed similarly to the non-stoquastic case considered in Ref. \cite{Barends2016_SM}. While the sequence represented in Fig. \ref{figS:digital_XYZ} will work perfectly in the $N_{\mathrm{Tr}} \rightarrow \infty$ limit, we note that the order of the unitaries $\{ \mathcal{S} \} = \{ U_{\mathrm{XY}}, U_{\mathrm{XZ}}, U_{\mathrm{YZ}}, U_{\mathrm{Z}} \}$, which form a Trotter step, will alter the final infidelity for the annealed state. Thus, for the digital simulation procedure we consider 24 permutations of unitaries for the set $\mathcal{S}$, and choose the sequence of the step which yields minimal infidelity.

We note that alternatively the XYZ model can be simulated with controlled-phase (ZZ) gates as described by Barends et al.  \cite{Barends2016_SM}.


\end{document}